\documentclass[journal]{IEEEtran}

\usepackage[table]{xcolor} % Must be loaded before other table packages
\usepackage{booktabs}
\usepackage{tabularx}
\usepackage{colortbl}

\usepackage[hidelinks]{hyperref}

\colorlet{shadecolor}{yellow}
\usepackage[pdftex]{graphicx}
\graphicspath{{../pdf/}{../jpeg/}}
\DeclareGraphicsExtensions{.pdf,.jpeg,.png}

\usepackage[cmex10]{amsmath}
\usepackage{array}
\usepackage{subfig}
\usepackage{mdwmath}
\usepackage{mdwtab}
\usepackage{eqparbox}
\usepackage{url}
\usepackage{amsfonts}
\usepackage{booktabs}
\usepackage{multirow}
\usepackage{makecell}
\usepackage{bbm}
\usepackage{stfloats}
\usepackage{caption}
\usepackage{stfloats}
\usepackage{array}

\begin{document}

\title{Silent Speech Interfaces in the Era of Large Language Models: A Comprehensive Taxonomy and Systematic Review}

\author{Kele Xu, Yifan Wang, Ming Feng, Qisheng Xu, Wuyang Chen, Yutao Dou, Cheng Yang, Huaimin Wang
        % <-this % stops a space
%\thanks{This paper was produced by the IEEE Publication Technology Group. They are in Piscataway, NJ.}% <-this % stops a space
\thanks{Kele Xu, Yifan Wang, Ming Feng, Qisheng Xu, Cheng Yang are with National Key Laboratory of Parallel and Distributed Computing, College of Computer Science and Technology, National University of Defense Technology. They are also with the State Key Laboratory of Complex \& Critical Software Environment, National University of Defense Technology, Changsha 410073, China.}
\thanks{Wuyang Chen is with College of Computer Science, Hunan Normal University, Changsha, China.}
\thanks{Yutao Dou is with College of Computer Science, Hunan University, Changsha, China.}
}

% \author{Anonymous}

% The paper headers
\markboth{
}{Kele Xu \MakeLowercase{\textit{et al.}}: Silent Speech Interface}

\maketitle

\begin{abstract}
Human-computer interaction has traditionally relied on the acoustic channel, a dependency that introduces systemic vulnerabilities to environmental noise, privacy constraints, and physiological speech impairments. Silent Speech Interfaces (SSIs) emerge as a transformative paradigm that bypasses the acoustic stage by decoding linguistic intent directly from the neuro-muscular-articulatory continuum. This review provides a high-level synthesis of the SSI landscape, transitioning from traditional transducer-centric analysis to a holistic intent-to-execution taxonomy. We systematically evaluate sensing modalities across four critical physiological interception points: neural oscillations, neuromuscular activation, articulatory kinematics (ultrasound/magnetometry), and pervasive active probing via acoustic or radio-frequency sensing.
Critically, we analyze the current paradigm shift from heuristic signal processing to Latent Semantic Alignment. In this new era, Large Language Models (LLMs) and deep generative architectures serve as high-level linguistic priors to resolve the ``informational sparsity'' and non-stationarity of biosignals. By mapping fragmented physiological gestures into structured semantic latent spaces, modern SSI frameworks have, for the first time, approached the Word Error Rate usability threshold required for real-world deployment. We further examine the transition of SSIs from bulky laboratory instrumentation to ``invisible interfaces'' integrated into commodity-grade wearables, such as earables and smart glasses. Finally, we outline a strategic roadmap addressing the ``user-dependency paradox'' through self-supervised foundation models and define the ethical boundaries of ``neuro-security'' to protect cognitive liberty in an increasingly interfaced world.
\end{abstract}


\IEEEpeerreviewmaketitle

\section{Introduction}
\label{sec:intro}

Human-computer interaction (HCI) has traditionally been predicated on the acoustic channel. Conventional Automatic Speech Recognition (ASR) systems-evolving from classical stochastic modeling to contemporary end-to-end neural architectures-operate under the prerequisite of high-fidelity audio acquisition \cite{rabiner2002tutorial, vaswani2017attention}. However, this fundamental reliance on airborne sound introduces systemic vulnerabilities: 
(i) \textbf{Environmental Fragility:} Performance degrades sharply in non-stationary noise or high-reverberation settings where the signal-to-noise ratio (SNR) is severely compromised \cite{lippmann1997speech}; 
(ii) \textbf{Social and Privacy Constraints:} The necessity of audible vocalization leads to sensitive information leakage and ``social friction'' in public or shared spaces \cite{edu2020smart}; and 
(iii) \textbf{Inclusivity Barriers:} Acoustic-dependent systems inherently exclude populations with severe phonation disorders, such as those suffering from laryngectomy-induced aphonia or neurodegenerative speech impairments \cite{beukelman2020augmentative}.

\begin{figure}[t]
\centering
\includegraphics[width=1.0\linewidth]{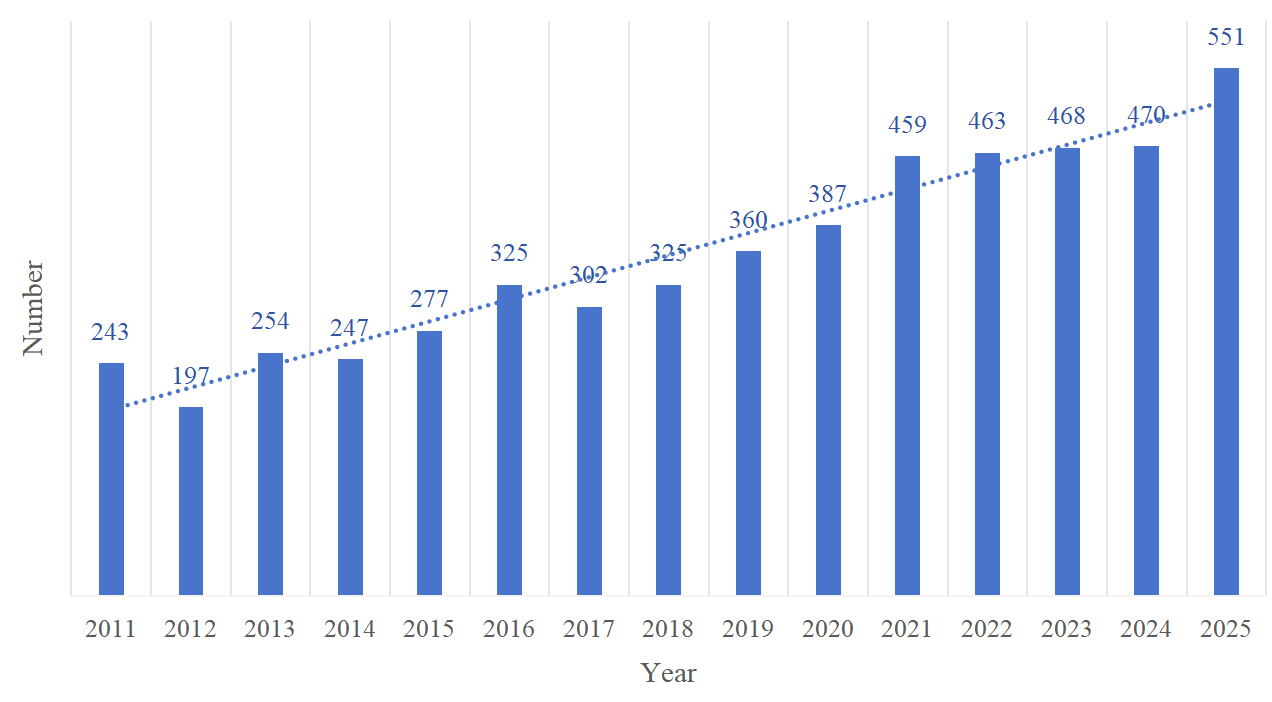} % 请确保路径正确
\caption{Total number of publications in the field of Silent Speech Interface (SSI). The statistics were retrieved from the Web of Science using the themes ``Silent speech'', ``Silent speech interface'', ``SSI'', ``Silent speech recognition'', ``Subvocal speech'', ``Mute speech recognition'', ``Non-audible speech'', ``Articulatory speech recognition'', ``Articulatory-to-text'', and ``Speech without sound'', covering the period from 2011 to 2025.}
\label{fig:publication}
\end{figure}

Silent Speech Interfaces (SSIs) emerge as a transformative paradigm that bypasses the acoustic channel by decoding linguistic intent directly from the neuro-muscular-articulatory chain of speech production \cite{denby2010silent,ren2025introduction,gonzalez2020silent,freitas2017introduction,zhang2025perspective}. 
To further illustrate the rapid growth and increasing research attention in this area, 
As shown in Fig.~\ref{fig:publication}, the number of annual publications related to SSI has risen consistently between 2011 and 2025. The speech production process originates from cortical motor intent, propagates via efferent neuromuscular activation, and culminates in the physical kinematics of the vocal tract (e.g., the tongue, lips, and velum). By intercepting this chain at distinct physiological stages---spanning neural oscillations (ECoG/EEG), myoelectric excitation (sEMG), and articulatory trajectories (ultrasound, magnetometry, or computer vision)-SSIs effectively decouple communicative intent from audible pressure waves. This facilitates a direct mapping from physiological precursors to semantic or synthesized acoustic representations, effectively redefining the boundaries of speech communication \cite{freitas2017introduction}.

\begin{figure}[htbp]
    \centering
    \includegraphics[width=0.99\linewidth]{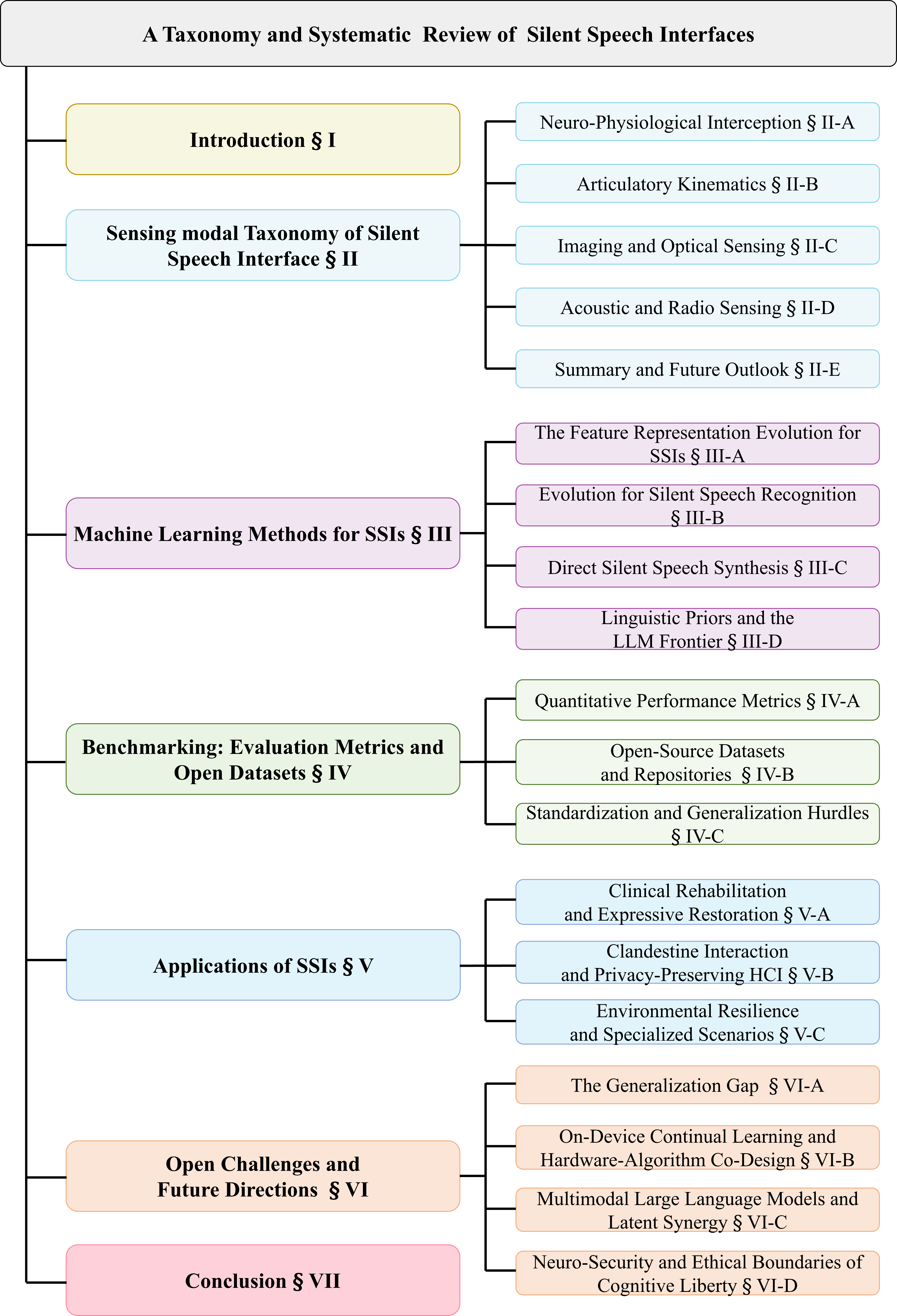}
    \caption{Framework of this survey paper.}
    \label{fig:framework}
\end{figure}

The current momentum in SSI research is propelled by three converging technological and societal imperatives: (i) \textbf{High-Performance Neuroprosthetics:} For individuals with profound speech loss, SSIs serve as a critical bridge to reclaiming communicative identity. Recent breakthroughs in high-density neural decoding have demonstrated the feasibility of synthesizing naturalistic speech from cortical activity with word error rates (WER) increasingly competitive with human-level performance \cite{willett2023high, zhou2025mindspeak}.
(ii) \textbf{Privacy-Preserving Ubiquity:} The proliferation of pervasive AI has intensified the demand for discreet interaction modalities. SSIs enable `unvoiced' commands, allowing users to engage with digital agents in silence, thereby mitigating the risk of acoustic eavesdropping in sensitive environments \cite{kapur2018alterego}.
(iii) \textbf{Acoustic Immunity and Robustness:} By operating in the non-acoustic domain, SSIs offer an intrinsic resolution to the ``cocktail party problem.'' Modalities such as sEMG and active acoustic sensing (AAS) maintain signal integrity irrespective of ambient sound pressure levels, ensuring deterministic performance in extreme industrial or aeronautical contexts \cite{jin2022earcommand, zhang2023echospeech}.

The technological landscape of SSIs has undergone a profound metamorphosis over the past decade, transitioning from bulky, laboratory-bound instrumentation to unobtrusive, commodity-grade wearables. Early research predominantly focused on high-precision but restrictive modalities, such as Ultrasound Tongue Imaging (UTI) \cite{kimura2019sottovoce} or Permanent Magnetic Articulography (PMA) \cite{gonzalez2016silent}. However, recent innovations in active sensing and flexible electronics have catalyzed a shift toward consumer-centric form factors. State-of-the-art SSIs now leverage transducers integrated into standard earbuds \cite{jin2022earcommand}, smart glasses \cite{zhang2023echospeech}, and wrist-worn devices \cite{zhang2024lipwatch} to track articulatory dynamics via acoustic echoes or inertial measurement units (IMUs). Notably, the emergence of contactless magnetometry and bio-impedance sensing has introduced high-sensitivity alternatives that circumvent the skin-coupling and motion-artifact challenges inherent in traditional sEMG.

In parallel with hardware evolution, the integration of Large Language Models (LLMs) and deep generative architectures has revolutionized the SSI computational backend. A historical bottleneck in SSI recognition has been the ``Silent Lombard Effect''-the systematic physiological variance in articulation that occurs in the absence of auditory feedback \cite{gaddy2020digital}. Contemporary frameworks address this via multi-modal contrastive learning and LLM-conditioned rescoring. By aligning fragmented physiological features with high-level linguistic priors in a unified latent space, these systems can robustly reconstruct missing articulatory information, achieving unprecedented accuracy in open-vocabulary, continuous speech tasks \cite{benster2024cross, su2025systematic}.

Despite these rapid advancements, the field lacks a comprehensive synthesis that bridges the gap between heterogeneous sensing modalities and the latest generative AI paradigms. This work provides a systematic review of the SSI state-of-the-art. We establish a technical taxonomy based on the speech production chain and critically evaluate the transition from traditional classification to modern Transformer-based and diffusion-based architectures. Furthermore, we explore emerging interaction paradigms in SSIs \cite{min2025exploring}, while addressing open challenges such as signal non-stationarity, cross-subject domain adaptation, and the ethical implications of ``thought-to-speech'' neurotechnology.

The specific contributions of this review are as follows (as shown in Fig \ref{fig:framework}):
\begin{itemize} 
    \item \textbf{Unified Physiological Taxonomy:} We provide a rigorous classification of SSI modalities based on their specific interception points within the neuro-muscular-articulatory chain, offering a comparative analysis of their physical constraints and information capacity.
    \item \textbf{Algorithmic Evolution Analysis:} We synthesize the transition from heuristic feature engineering to deep representation learning, with a specific focus on cross-modal latent alignment, self-supervised pre-training, and LLM-driven decoding frameworks.
    \item \textbf{Multifaceted Application Scenarios:} We delineate the diverse operational landscapes of SSIs, ranging from clinical restorative neuroprosthetics for speech disorders to secure communication in public spaces and high-noise industrial or aerospace contexts.
    \item \textbf{Benchmarking and Open Science:} We provide a curated repository of open-source datasets and standardized evaluation metrics to facilitate reproducible research and cross-modal benchmarking.
    \item \textbf{Challenges and Research Roadmap:} We identify critical gaps in subject-independent generalization and zero-shot transfer, proposing a roadmap for the next generation of robust, ethically-aligned silent speech technologies.
\end{itemize}

\section{Sensing modal Taxonomy of Silent Speech Interface}\label{sec:modalities}

\begin{figure*}[t]
\centering
\includegraphics[width=1.0\linewidth]{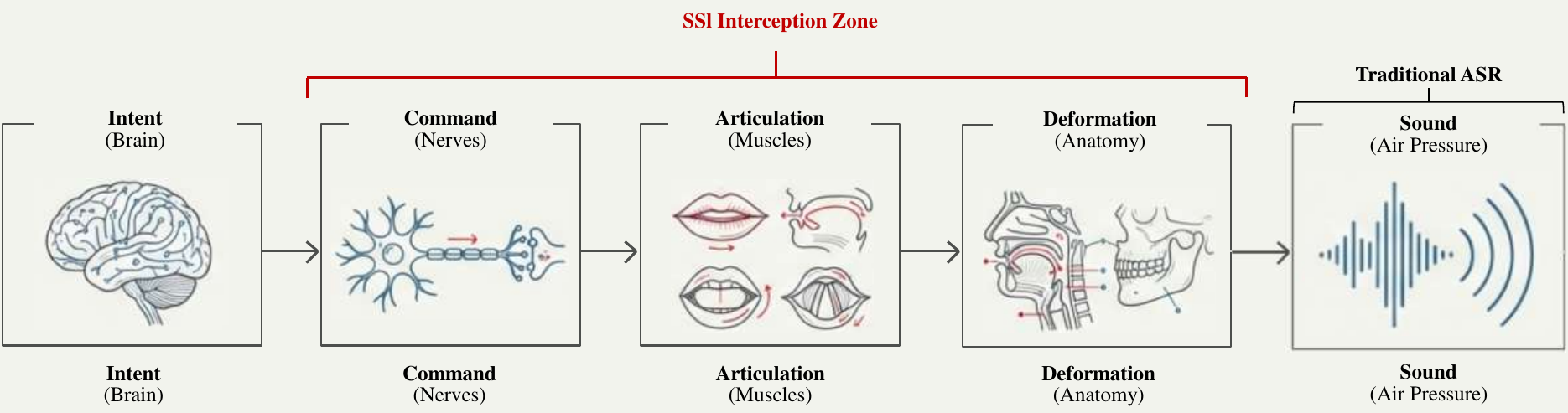} % 请确保路径正确
\caption{The The Neuro-Muscular-Articulatory (NMA) Continuum and SSI Interception Boundaries. The schematic delineates the physiological stages of speech production, from cortical intent and neural signaling to muscular activation and anatomical deformation. The SSI Interception Zone (highlighted in red) defines the functional domain of technologies that bypass the acoustic stage (Traditional ASR) to restore or augment communicative agency via non-vocal biosignals.}
\label{fig:chain_speech}
\end{figure*}

The operational efficacy of a SSI is intrinsically determined by its point of interception along the neuro-muscular-articulatory (NMA) continuum~\cite{ephratt2011linguistic}. As conceptualized in Fig. \ref{fig:chain_speech}, the SSI interception zone encompasses the entire non-acoustic trajectory of speech production, effectively circumventing the acoustic channel to decouple linguistic intent from atmospheric pressure waves~\cite{assouar2018acoustic}.

These modalities form a hierarchical taxonomy-systematically categorized in Table I-based on the depth of the physical phenomena they transduce, ranging from proximal cortical oscillations to distal kinematic execution~\cite{guo2025cross,kaur2026convbi}. This hierarchy illustrates a fundamental technical tension: the trade-off between the high degrees-of-freedom ($DoF$) and semantic density of proximal neural decoders versus the superior social acceptability of distal wearable sensors~\cite{hongsoft,chowdhury2025summary}. The transition from these physiological stages to a functional interface requires a diverse array of transduction technologies, each facing unique constraints such as the ``observability–pervasiveness paradox.''

\subsection{Neuro-Physiological Interception}

Neuro-physiological sensing captures speech intent at its biological source, bypassing the physical constraints of the vocal tract. These modalities must contend with low signal-to-noise ratios ($SNR$) and high inter-subject variability within the neural manifold.

\subsubsection{Cortical Intent}
Cortical sensing seeks to decode imagined or intended speech by monitoring brain activity, navigating a fundamental tension between non-invasive scalability and invasive fidelity. Scalp-recorded Electroencephalography (EEG) remains attractive due to its accessibility and high temporal resolution~\cite{krishna2020continuous,kim2022ultrathin}; however, spatial information is severely attenuated by the cranial volume conductor, resulting in blurred cortical representations. Recent work has therefore shifted toward \textit{Neural--Linguistic Alignment}~\cite{zhou2025mindspeak}, where self-supervised pre-training maps noisy neural embeddings onto structured linguistic latent spaces, leveraging semantic priors to regularize decoding.

In contrast, Electrocorticography (ECoG) provides direct access to High Gamma Energy (HGE, 70--170~Hz), which correlates strongly with articulatory kinematics~\cite{chen2025decoding,gonzalez2020silent}. Beyond its superior spectral fidelity, ECoG offers notable longitudinal stability~\cite{wadkins2019continuous,luo2025self,freitas2017introduction,stavisky2025restoring}; unlike penetrating microelectrodes, subdural grids mitigate gliosis, supporting durable chronic recordings~\cite{card2025long}. Consequently, ECoG has emerged as the gold standard for continuous speech reconstruction in restorative neuroprosthetic systems~\cite{freitas2017introduction,krishna2020continuous,brumberg2010brain,willett2023high}.

\subsubsection{Neuromuscular Execution}Surface Electromyography (sEMG) intercepts signals at the neuromuscular junction~\cite{fitriah2022eeg,mohanchandra2016communication}. While neural decoders grapple with abstract cortical intent, sEMG captures physical muscle activation. A critical advantage is the \textit{electromechanical delay}-a 60 ms window where electrical excitation precedes physical articulation~\cite{krishna2020continuous,mohanchandra2016communication}. This pre-emptive signal facilitates low-latency predictive decoding essential for real-time applications~\cite{su2025systematic}. The transition toward epidermal electronics and textile sensors~\cite{tang2025textile} has addressed historical limitations regarding motion artifacts and skin-impedance instability.

\subsection{Articulatory Kinematics}Articulatory kinematics measure the mechanical execution of the vocal tract, offering a high-fidelity proxy for phonetic output robust to electromagnetic interference.

\subsubsection{The Magnetic Frontier}Electromagnetic Articulography (EMA) has served as the phonological benchmark~\cite{pattem2018optimal,bonaventura2013assessment}, providing sub-millimeter 3D tracking~\cite{gonzalez2016silent,rebernik2021review}. However, tethered intra-oral coils restrict EMA to laboratory use. Contemporary shifts toward \textit{Permanent Magnetic Articulography (PMA)} reflect a drive toward untethered sensing. By utilizing high-sensitivity Tunneling Magnetoresistance (TMR) sensors in wearable headsets~\cite{srivastava2024whispering}, PMA decouples articulatory motion from ambient noise~\cite{srivastava2024whispering,menezes2025multimodal}. Furthermore, Optically Pumped Magnetometers (OPMs) enable contactless sensing of deep musculature inaccessible to surface electrodes~\cite{dash2025silent}.

\subsubsection{Anatomical Synergy}
A significant trend is the development of invisible interfaces exploiting anatomical coupling, such as the temporomandibular joint (TMJ) bridge between articulators and the ear canal~\cite{pizolato2011speech}. Systems leveraging \textit{Ear Canal Deformation (ECDM)} repurpose ubiquitous earables for silent speech decoding~\cite{zhou2025baroauth,wang2021eardynamic}. While facing informational sparsity, their form-factor makes them candidates for consumer-grade SSI. Current research addresses this through \textit{Vibro-Acoustic Fusion}, pairing proximity sensors with bone-conducted inertial measurement units (IMUs) to filter artifacts~\cite{dai2025poster}.

\newcolumntype{Y}[1]{>{\raggedright\arraybackslash\hsize=#1\hsize}X}

\begin{table*}[htbp]
\centering
\caption{Taxonomy and Comparative Analysis of Silent Speech Sensing Modalities}
\label{tab:ssi_modalities_refined}
\footnotesize % Standard for dense Transaction tables to maintain clarity
\renewcommand{\arraystretch}{1.1} % Compact but readable
\setlength{\tabcolsep}{4pt}      % Optimized horizontal spacing

% Adjusting column weights: Sensing Principle (0.8), Advantages (1.1), Bottlenecks (1.1)
\begin{tabularx}{\textwidth}{@{} l l Y{0.8} Y{1.1} Y{1.1} @{}}
\toprule
\textbf{Category} & \textbf{Modality} & \textbf{Sensing Principle} & \textbf{Advantages \& Capacity} & \textbf{Technical Bottlenecks} \\ \midrule

\multicolumn{5}{l}{\textit{Neuro-Physiological (Proximal)}} \\
Neuro-Physio. & EEG & Scalp potentials & Non-invasive; cortical intent. & Low SNR; cranial attenuation. \\
& ECoG & Sub-dural oscil. & High-fidelity HGE access; stable. & Surgical invasiveness; accessibility. \\
& sEMG & Neuromuscular & Ultra-low latency; pre-emptive. & Electrode drift; motion artifacts. \\ \midrule

\multicolumn{5}{l}{\textit{Articulatory Kinematics (Distal)}} \\
Kinematic & PMA / EMA & Magnetic tracking & Sub-mm precision; gold standard. & Wired constraints; interference. \\
& Earables & Proximity / Sonar & High social acceptance; pervasive. & Informational sparsity; drift. \\
& Intra-oral & Electro-optical & 3D oral profiling; vowel distinction. & High obtrusiveness; ergonomics. \\ \midrule

\multicolumn{5}{l}{\textit{Imaging and Optical Sensing}} \\
Imaging & rtMRI & Magnetic res. & Full vocal tract geometry; nasality. & Non-portable; high cost. \\
& UTI & Submental US & Visualizes lingual dynamics; safe. & Acoustic shadows; misalignment. \\
& VSR / Depth & Optical / RGB-D & Ubiquitous hardware; intuitive. & Light sensitivity; privacy. \\ \midrule

 \multicolumn{5}{l}{\textit{Acoustic and Radio Frequency (RF)}} \\
Acoustic / RF & Active Sonar & FMCW / Pulse & Noise immunity; micron-tracking. & Multi-path interference. \\
& NAM / IMU & Tissue cond. & Robust in noise; captures sub-voice. & Low-pass filtering effect. \\
& mmWave & Micro-Doppler & \textbf{Contactless}; mask-penetrable. & Sensitive to non-speech motion. \\ \bottomrule

\multicolumn{5}{@{}p{\textwidth}@{}}{\scriptsize \textit{Note: SNR: Signal-to-Noise Ratio; HGE: High Gamma Energy; FMCW: Frequency-Modulated Continuous Wave. Bold entries denote state-of-the-art (SOTA) frontiers.}} \\
\end{tabularx}
\end{table*}

\subsubsection{Intra-Oral Sensing}
Intra-oral sensing remains indispensable for maximum phonetic coverage, transitioning from capacitive Electropalatography (EPG)~\cite{mat2021technology,hardcastle1997electropalatography} to hybrid \textit{Electro-Optical Stomatography (EOS)}~\cite{preuss2015optical,stone2020cross} to resolve vowel-blindness. Traditional EPG is restricted to contact dynamics; by integrating optical distance sensors into a pseudopalate, EOS provides a 3D oral profile. These architectures are essential for phoneme-to-text systems requiring high-resolution distinction between consonants and vowels, serving as critical clinical tools where earable fidelity is insufficient.

\subsection{Imaging and Optical Sensing}
Imaging modalities offer direct visualization of vocal tract geometry, providing immunity to electromagnetic interference while confronting an inherent \textit{observability--pervasiveness paradox}.

\subsubsection{Ultrasound Tongue Imaging (UTI)}
UTI employs B-mode ultrasound (2--10~MHz) to visualize internal lingual dynamics~\cite{ji2018updating,li2019denoising}. A submental transducer emits acoustic waves that reflect at the tongue--air interface, producing a characteristic bright contour representing the midsagittal tongue shape~\cite{stone2005guide,xia2024systematic,xu2016comparative}. Methodologically, the field has progressed from linear PCA-based representations toward \textit{Generative Neural Manifolds}. Spatial Transformer Networks (STNs) are increasingly used to correct probe misalignment through real-time affine adaptation~\cite{toth2023adaptation}, while Denoising Convolutional Autoencoders (DCAEs) mitigate acoustic shadows induced by jaw occlusion during complex alveolar articulations~\cite{li2019denoising,xu2019ultrasound}.

\subsubsection{Real-time MRI (rtMRI)}
Real-time MRI provides a comprehensive view of vocal tract dynamics, including critical structures such as the velum, thereby enabling principled modeling of nasality and coarticulation beyond acoustics~\cite{toutios2016advances,asadiabadi2020vocal,somandepalli2017semantic}. Although constrained by limited temporal resolution (typically 5--50~Hz), rtMRI has increasingly assumed the role of a \textit{teacher model} within knowledge distillation frameworks. High-resolution articulatory sequences supervise deep networks whose learned anatomical representations can be transferred to lightweight \textit{student} sensors (e.g., earables), facilitating practical SSI deployment under reduced observability.

\subsubsection{Video Lip Reading}
Visual Speech Recognition (VSR) emulates human lip-reading by exploiting commodity cameras to capture visible articulatory cues~\cite{li2024ai,chen2024comprehensive}. Feature representations range from shape-based geometric descriptors and appearance-based region-of-interest (ROI) pixels to deep spatio-temporal models (e.g., CNN--LSTM). Despite its intuitive appeal, VSR is vulnerable to illumination variability and privacy concerns. This has motivated a transition toward \textit{Depth Sensing} and wearable \textit{Infrared (IR)} systems~\cite{igarashi2024silent}, which infer facial geometry or submental motion without exposing full-frame visual content.

\subsection{Acoustic and Radio Sensing}
Acoustic and Radio Frequency (RF) sensing treat the vocal tract as a dynamic, reflective environment.

\subsubsection{Active Sonar and RF Micro-Doppler}

Active Acoustic Sensing repurposes hardware into sonar systems, emitting near-ultrasound pulses (16--22 kHz) to track micron-level skin deformations~\cite{luo2021end}. Advancements in time-shifted FMCW carriers have addressed multi-user interference, enabling device-free SSI in shared spaces~\cite{zhou2025m2silent}. Similarly, millimeter-wave (mmWave) radar (77--120 GHz) detects \textit{micro-Doppler signatures} of articulatory shaping through clothing or masks. Challenges in spatio-temporal disambiguation are addressed through Stepped Frequency Continuous Wave (SFCW) systems capturing vocal tract transmission spectra~\cite{wagner2022silent}.

\subsubsection{Internal Conducted Signals}

Non-Audible Murmur (NAM) microphones capture speech at the point of origin but suffer from a low-pass filter effect of tissue. Breakthroughs involve \textit{Cross-Modal Distillation}, where teacher models trained on acoustic or rtMRI data guide student models processing NAM or IMU signals. This reconstructs missing high-frequency phonetic latent spaces, enabling user-independent robustness even when sensor placement varies~\cite{srivastava2024whispering}.

\subsection{Summary and Future Outlook}
The evolution of SSI research reflects a shift from isolated modality exploration toward \textit{synergistic multi-modal fusion}. Each sensing paradigm seeks an operating point within the observability--pervasiveness trade-off. Future progress hinges on cross-modal knowledge distillation and edge-efficient learning, enabling gold-standard laboratory signals to inform privacy-aware, robust, and socially acceptable communication interfaces.

\newcolumntype{L}{>{\raggedright\arraybackslash}X}
\newcolumntype{P}[1]{>{\raggedright\arraybackslash}p{#1}}
% \begin{table*}[htbp]
\begin{table*}[!t]
\centering
\caption{Comprehensive Synthesis of ML Methodologies and Performance Benchmarks Across the NMA Continuum}
\tiny
\label{table:unified_ssr_benchmarks}
\footnotesize 

\renewcommand{\arraystretch}{1.1} 
\setlength{\tabcolsep}{2.5pt}      
\begin{tabularx}{\textwidth}{@{} P{1.6cm} c L P{1.6cm} P{1.4cm} P{1.8cm} L l @{}}
% \begin{tabularx}{\textwidth}{@{} l c l l l c l l @{}}
\toprule
\textbf{Modality} & \textbf{Architecture } & \textbf{Dataset / Corpus} & \textbf{Vocab.} & \textbf{Metric} & \textbf{Result} & \textbf{Population} & \textbf{Ref.} \\ \midrule
\multicolumn{8}{@{}l@{}}{\textit{Neuro-Physiological}} \\
% \textit{Neuro-Physiological} & & & & & & & \\
Utah Array & Trans-based & BrainGate2 (T15) & 125k & W. Acc. & 99.2\% & ALS / Dysarthria & \cite{card2025long} \\
Utah Array & Deep Neural Network & Sentence Decod. & 125k & Acc. & 97.5\% & Tetraplegia / ALS & \cite{khan2025invasive} \\
Utah Array & RNN & BrainGate2 (T12) & 125k & Acc. & 96.5\% & ALS / Dysarthria & \cite{card2025long} \\
ECoG & Seq2Seq + LLM & Kinematic Decod. & Cont. & WER & 3.0\% & Healthy / Clinical & \cite{gonzalez2020silent} \\
ECoG & CNN (InceptionTime) & Clinical (Corticom) & 14 Cmds & M. Acc. & 97.1\% & ALS Participant & \cite{luo2025self} \\
EEG / EMG & Conformer + HTNet & Japanese Corpus & 64 Words & W. Acc. & 95.3\% & Clinical Patients & \cite{inoue2025silent}  \\
EEG & Stacking Classifier & KaraOne / FEIS & 4--16 & Acc. & 75--86.2\% & Communication Dis. & \cite{ramkumar2023aa} \\
EEG & Large Brain Language Model & 12 Word Set & 12 Words & Acc. & 77.3\% & Healthy Subjects & \cite{zhou2025mindspeak} \\
EEG & LSTM & SSDC & 6 silent commands & Acc., F1-score & 22.79\% & 10 participants & \cite{lotey2025native} \\
sEMG & MTL-Transformer & HD-sEMG Corpus & 11 Cmds & Acc. & 98.7\% & Healthy / Prosthetic & \cite{xie2025neural} \\
sEMG & DNN-HMM Hybrid & WSJ0 5k Corpus & 5,000 & WER & 6.45\% & Healthy Subjects & \cite{chowdhury2025decoding} \\
sEMG & HMM-SGMM (fMLLR) & Laryngectomy & 2.5k & WER & 8.9--13.6\% & Laryngectomees & \cite{meltzner2017silent} \\ 
sEMG & DNN+LDA & Custom Dataset & 5k Utter. & Objective & - & Healthy Subjects & \cite{janke2017emg} \\
sEMG & ResNet+Transformer & Custom Dataset & - & WER & 0.2696 & - & \cite{dai2025novel} \\
sEMG & DNN + LDA & Custom Dataset & 5k Utter. & Objective & - & Healthy Subjects & \cite{janke2017emg} \\
sEMG & SVM, KNN and DTW & Custom Dataset & 10 digits & WER & 0.343 & 3 speakers & \cite{freitas2014enhancing} \\
sEMG & DT & EMG-UKA Trial Corpus & 50 words & W.~Acc & 95.55\% & 4 speakers & \cite{abdullah2020computationally} \\
\midrule

 \multicolumn{8}{@{}l@{}}{\textit{Articulatory (UTI)}} \\
% \textit{Articulatory (UTI)} & & & & & & & \\
UTI (Submental) & Conformer + bi-LSTM & TaL80 / TaL Corpus & 1k Utter. & MCD & 3.03 dB & Multi-speaker & \cite{ibrahimov2025conformer} \\
UTI & ViT-based & Children's SSD Set & Sent. & Acc. & 88.9\% & Children (SSD) & \cite{xia2024systematic} \\
UTI & Diffusion/ GAN & TaL80 / Mandarin & Utter. & MCD & 5.21 dB & Healthy Subjects & \cite{zheng2024incorporating} \\
UTI & DNN-HMM with DCAE & 2010 Challenge & - & WER & 6.17\% & Healthy Subjects & \cite{li2019denoising} \\
UTI & DNN + HMM & 1023 Word Set & 1,023 & W. Acc. & 94.1\% & Healthy Subjects & \cite{stone2021silent} \\ 
UTI & PCA + MLP & IEEE/Harvard & - & NMSE (LSF)  & 11–16\%  & Healthy Subjects  & \cite{hueber2007eigentongue} \\
UTI  & ViT-based Self-supervised  & UXTD &  4609 Sent. & Acc. & 13.33\% & Healthy Subjects
& \cite{you2023raw} \\
UTI  & 3DCNN Self-supervised & WSJ0 / TJU UTI & 735k / 147k  & MSE & 21.7-32.6 & Healthy Subjects & \cite{wu2018predicting} \\
UTI & HMM LM (DCT-based) & Visual TIMIT / WSJ0 & 5k / 20k & W. Acc & 69–79\% & Healthy Subject & \cite{cai2011recognition} \\
UTI & Sequential CAE  & 2010 Challenge & - & WER & 5.96\% & Healthy Subjects 
& \cite{xu2019ultrasound} \\
UTI & KNN  & Custom Dataset & 30 German words &  Accuracy & 52\% & Single speaker 
& \cite{stone2016silent} \\
UTI & Trans-based  & WSJ0 2k Corpus & 2000 &  CER, WER & CER = 10.1\%, WER = 20.5\% & Single speaker 
& \cite{kimura2020end} \\
UTI & AutoEncoder  & Custom Dataset & Sentences &  NMSE & - & Single speaker 
& \cite{gosztolya2019autoencoder} \\

\midrule

\multicolumn{8}{@{}l@{}}{\textit{Kinematic \& Other}} \\
% \textit{Kinematic \& Other} & & & & & & & \\
Earables (MxG) & Deep Learning / Ridge Classifier & Sonera Custom & 12 Cmds & Acc. & 86.5\% & Wearable Users & \cite{dash2025silent} \\
EMA & BLSTM-HMM / Deep CCA & 132 Phrases & 39 Phon. & PER & 33.9\% & Laryngectomees & \cite{kim2017speaker} \\
EMA & GMM-HMM & Custom Dataset & 37 Phon. & PER & 47.5\% & 5 native English speakers & \cite{wang2015speaker} \\
Video (Lip) & LipNet (STCNN + Bi-GRU) & GRID Corpus & 51 Words & S. Acc. & 95.2\% & Healthy Subjects & \cite{lee2021biosignal} \\
Video (Lip) & VALLR (Video-Trans + LLM) & LRS3 Dataset & Large & WER & 18.7\% & Healthy (Data-eff.) & \cite{su2025multimodal} \\ \midrule

\multicolumn{8}{@{}l@{}}{\textit{Acoustic \& RF}} \\
% \textit{Acoustic \& RF} & & & & & & & \\
SAAS (PMUT) & ResNet (Residual Network) & Daily Comm. & 10 Sent. & Acc. & 99.8\% & Potential Patients & \cite{liu2025machine} \\
mmWave & mSilent (CNN + Trans. S2S) & mSilent Corpus & 1k Sent. & WER & 9.5\% & Healthy (Non-contact) & \cite{zeng2023msilent} \\
Airborne Ultrasound & CNN + CTC & Custom Dataset & 31 Cmds & WER & 4.5\% & Healthy Subjects & \cite{zhang2023echospeech} \\
In-ear US & EarS SR (Hierarchical CNN) & Custom Dataset & 50 Words & Acc. & 93.0\% & Wearable (In-ear) & \cite{sun2024earssr} \\
In-ear Sonar & EarCommand (CRNN + CTC) & EarCommand & 57 Cmds & WER & 10.0\% & Healthy Subjects & \cite{jin2022earcommand} \\
Ingressive Air & Trained Recognizer & SilentVoice & 80 Sent. & WER & 5.0--6.7\% & Healthy Subjects & \cite{fukumoto2018silentvoice} \\ \bottomrule
\end{tabularx}
\end{table*}

\section{Machine Learning Methods for SSIs}\label{sec:algorithms}

The computational core of modern SSIs involves the high-fidelity transformation of non-stationary biosignals $\mathbf{X} \in \mathbb{R}^{C \times T}$ into discrete linguistic tokens or continuous acoustic manifolds. As the field matures, the primary research challenge has transitioned from heuristic pattern recognition to end-to-end (E2E) neural latent alignment~\cite{taylor2019analysis,prabhavalkar2023end}. This trajectory focuses on resolving the spatio-temporal friction between fragmented physiological gestures and sequential semantic units, reflecting a broader transition from hand-crafted heuristics toward representation learning optimized for cross-modal invariance.

\subsection{The Feature Representation Evolution for SSIs}
The efficacy of the decoding backend is fundamentally contingent upon extracting articulatory embeddings that remain robust against environmental non-stationarity and inter-subject anatomical variability (see Table \ref{table:unified_ssr_benchmarks}).

\subsubsection{Heuristic Foundations and Classical Statistical Manifolds}

In the foundational era of SSI research, feature extraction focused on capturing the biomechanical and statistical properties of articulation through hand-crafted descriptors derived from predefined rules. For surface electromyography (sEMG), Time-Domain (TD) features-including Root Mean Square (RMS), Mean Absolute Value (MAV)~\cite{khushaba2020recursive}, and Waveform Length (WL)-remain standard as they directly quantify motor unit recruitment intensity without the computational latency of spectral transformation \cite{schultz2017biosignal}.  % chenwuyang:是survey，因此没有体现在表格中
For high-dimensional imaging modalities such as ultrasound and real-time MRI (rtMRI)~\cite{lim2024speech}, Principal Component Analysis (PCA) served as a pioneering technique to transform correlated variables into uncorrelated principal components~\cite{dawson2016methods}. % chenwuyang:是一篇纯分析文章，因此没有体现在表格中
This allowed researchers to project complex articulatory contours~\cite{asadiabadi2020vocal,lingala2017fast} into lower-dimensional subspaces, generating interpretable representations such as EigenTongues~\cite{hueber2007eigentongue} or EigenLips~\cite{bregler1994eigenlips}. % chenwuyang:\cite{asadiabadi2020vocal,lingala2017fast} 分别是纯运动分析和成像的文章，因此没有体现在表格中。 \cite{hueber2007eigentongue} 加在了表格中，但实际任务会有些区别属于mapping。 \cite{bregler1994eigenlips} 属于AVSR，不是SSI，因此没有体现在表格中。
Simultaneously, Gabor filter banks and wavelets were employed to extract local features by representing articulatory images as planar sinusoids~\cite{berry2010automatic}. % chenwuyang: 找不到这篇文章
As these methods matured, the Discrete Cosine Transform (DCT)~\cite{cai2011recognition} emerged as a superior alternative to PCA due to its enhanced energy compaction and computational efficiency. % chenwuyang: \cite{cai2011recognition}  加在了表格中

Beyond physical descriptors, classical statistical manifolds provided the primary framework for signal purification. Independent Component Analysis (ICA) became indispensable for blind source separation~\cite{jenson2014temporal}, % chenwuyang: \cite{jenson2014temporal}是neurophysiology方向的，因此没有体现在表格中
isolating speech-related neural intent from confounding biological artifacts in EEG and ECoG streams. Furthermore, Linear Discriminant Analysis (LDA) is frequently employed as a post-extraction refiner; by maximizing the ratio of between-class to within-class variance, LDA optimizes phonetic discriminability, ensuring that even low-resource models can resolve subtle articulatory contrasts \cite{janke2017emg}.% chenwuyang: \cite{janke2017emg}  加在了表格中

\subsubsection{Autonomous Representation and Deep Latent Learning}

The integration of Deep Learning (DL) has ushered in a ``feature-free'' paradigm, where neural networks internalize the feature extraction process by learning hierarchical representations directly from raw data. Deep Autoencoders (DAEs)~\cite{vachhani2017deep} became prominent for their ability to learn compact representations of unlabeled data, maintaining high visual fidelity even under significant dimensionality reduction~\cite{qiu2018denoising,li2019denoising}. Convolutional Neural Networks (CNNs) established a hierarchy of progressively higher-order features that outperformed traditional hand-crafted methods in pattern recognition.% \cite{li2019denoising} 已经有了

Specifically, 2D-CNNs effectively supersede traditional DCT/PCA by identifying localized articulatory edges and textural cues~\cite{ji2018updating}. To capture dynamic articulatory trajectories, 3D Convolutional Neural Networks (3DCNNs) capture motion by explicitly incorporating the time dimension through 3D kernels~\cite{wu2018predicting,angrick2019speech}.
This evolved into Sequential Convolutional Autoencoders (Sequential CAEs), which utilize LSTM models to process entire image sequences as fixed-length vectors~\cite{xu2019ultrasound}. To combat inherent imaging noise and speckle artifacts, Denoising Convolutional Autoencoders (DCAEs) were developed to preserve spatial information while filtering disturbances~\cite{lai2016deep}. % chenwuyang: \cite{xu2019ultrasound} 已加入表格

\subsubsection{Self-Supervised and Multimodal Learning}
Recent advancements have shifted toward overcoming the chronic scarcity of annotated data. Self-supervised learning (SSL) paradigms based on mask modeling allow models to infer high-dimensional abstract features directly from unlabeled data using Vision Transformer (ViT) architectures~\cite{you2023raw}. % chenwuyang:\cite{you2023raw} 已加入表格，实际是pretrain+finetune

Current state-of-the-art decoders leverage Self-Attention mechanisms within Transformer and Conformer architectures to resolve the complex temporal `smearing' inherent in co-articulation \cite{su2025systematic}. % chenwuyang: \cite{su2025systematic} 是survey，因此没有体现在表格中
Unlike traditional frame-level features, these models weight extracted embeddings based on their global linguistic importance, capturing the fluid dynamics of continuous speech production. Furthermore, multimodal approaches, such as FusionNet, enhance classification accuracy by combining raw imaging data with texture features extracted via Local Binary Patterns (LBP)~\cite{kabakoff2021extending}. 

\subsubsection{Physical Modeling and Biological Priors as Inductive Biases}
Despite the dominance of data-driven methods, modern SSI research has re-integrated physical modeling as a rigorous form of inductive bias~\cite{wilhelms1995physiological,sifakis2006simulating}. This hybrid approach is critical in clinical scenarios where data scarcity limits the generalization of black-box models~\cite{yang2025towards}. A prominent paradigm is Biomechanical Constraining~\cite{xu2016contour,xu2020predicting}, where Active Contour Models (Snakes) are embedded into neural pipelines to enforce anatomical plausibility~\cite{xu2016comparative}. %  % chenwuyang: \cite{xu2016comparative} 是分析性的文章，因此没有体现在表格中
By constraining tongue tracking within physiologically valid boundaries, these models prevent the `hallucination' of impossible articulatory states during decoding \cite{tang2025textile}. % chenwuyang: \cite{tang2025textile} 搜不到这篇文献

Furthermore, the synthesis of computer vision and musculoskeletal modeling has catalyzed the emergence of Markerless Kinematic Extraction. Frameworks such as DeepLabCut extract ``Virtual Markers'' from non-invasive video or ultrasound streams~\cite{wrench2022beyond,sun2025extraction}, simulating the precision of invasive electromagnetic midsagittal articulography (EMA) sensors without procedural discomfort. These kinematic features are inherently more invariant to individual physiological appearances than raw pixel or myoelectric data, facilitating superior cross-subject transfer learning and addressing the long-standing challenge of speaker-independent calibration \cite{srivastava2024whispering}. % chenwuyang：搜不到这篇文章

\subsection{Evolution for Silent Speech Recognition}
The algorithmic landscape of Silent Speech Recognition (SSR) has undergone a definitive transition from heuristic-based statistical modeling to high-capacity, end-to-end neural architectures~\cite{dai2025novel}. This evolution reflects the field's increasing capacity to resolve the nonlinear mapping between stochastic biosignals and discrete linguistic units. Modern decoding frameworks are characterized by their ability to model complex temporal dependencies, extract invariant spatial features, and mitigate the pervasive challenges of data scarcity and inter-subject anatomical variability.
% xuqisheng: wang2025ssr 这篇文献删除了，数据不涉及SSI   dai2025novel这篇论文表格没引用，已添加

\subsubsection{Statistical Foundations and Heuristic Classifiers} Early SSR systems were primarily anchored in classical statistical frameworks designed for isolated word recognition or constrained. Hidden Markov Models (HMM) historically served as the architectural backbone~\cite{juan2016current,ji2018updating,wang2015speaker},
% xuqisheng: juan2016current综述，不适合放到表格里  ji2018updating这篇属于UTI和Lip的多模态，暂时还未归类到表格中  wang2015speaker属于EMA，已添加至表格
often paired with Gaussian Mixture Models (GMM-HMM) to represent the spectral or myoelectric properties of articulatory states. While these models provided foundational benchmarks, they often exhibited high error rates due to their limited capacity to model the non-stationary dynamics of silent speech. Complementary techniques such as Dynamic Time Warping (DTW) were employed for template matching, particularly to compensate for temporal distortions in kinematic sensor data~\cite{freitas2014enhancing},
% xuqisheng: chorowski2015attention原来引用的这篇论文不属于SSI的范畴，输入直接就是语音信号了 freitas2014enhancing已添加至表格中
though their computational complexity restricted their application in continuous speech tasks~\cite{stone2016silent}. For low-resource scenarios, discriminative classifiers such as Support Vector Machines (SVM) and Random Forests remain relevant for identifying phonemic units from small-scale EEG or sEMG datasets where deep learning remains prone to overfitting.
For instance, TD6 time-domain features combined with a decision tree classifier were shown to achieve closed-set recognition of sEMG signals while reducing the number of channels from seven to five without significant performance degradation~\cite{abdullah2020computationally}.

\subsubsection{Neural Representation and Spatiotemporal Modeling} The integration of Deep Neural Networks (DNN) signaled a paradigm shift toward capturing the latent nonlinearities inherent in neuromuscular and imaging data. An initial milestone involved the DNN-HMM hybrid, which replaced GMMs with deep architectures for emission probability estimation, significantly reducing Word Error Rates (WER) in ultrasound-based systems~\cite{ji2018updating,kimura2020end}. % xuqisheng: kimura2020end属于UTI+Lip，已归类到表格中
To further exploit the spatial structures of high-density sEMG grids or visual visemes, CNN were deployed as hierarchical feature extractors. In particular, 3D-CNNs have become indispensable for video and ultrasound sequences, convolving over both spatial and temporal dimensions to capture the fluid dynamics of articulatory trajectories~\cite{xia2024systematic}. % xuqisheng:已加入表格
To model the sequential nature of speech and the complex effects of co-articulation, recurrent architectures such as Long Short-Term Memory (LSTM) and Bidirectional LSTMs (Bi-LSTM) have become standard, utilizing both past and future context to resolve phonetic ambiguity~\cite{gosztolya2019autoencoder}. % xuqisheng: 已加入表格

\subsubsection{End-to-End Alignment and Attention Mechanisms} State-of-the-art SSR decoders have largely moved toward End-to-End (E2E) learning frameworks that map raw biosignals directly to text without intermediate phonetic alignment. Connectionist Temporal Classification (CTC) has emerged as a critical loss function in this context, providing a mathematical solution for aligning unconstrained articulatory signals with target text sequences of varying lengths. This is frequently paired with Sequence-to-Sequence (Seq2Seq) encoder-decoder architectures that utilize Attention mechanisms to intelligently weight relevant articulatory frames during the decoding process~\cite{lotey2025native,zeng2023msilent}. % xuqisheng: 这两篇都已加入表格
Currently, the Conformer architecture-which synthesizes the local feature extraction of CNNs with the global context modeling of Transformers-represents the performance frontier, achieving superior results across diverse modalities by capturing both fine-grained kinematic details and high-level semantic dependencies~\cite{benster2024cross}.  % xuqisheng: 已加入表格

\subsubsection{Generalization and Semantic Refinement Strategies} 

To overcome the ``data bottleneck'' and anatomical variance among users, researchers employ sophisticated robustness strategies. Data Augmentation techniques, such as SpecAugment and temporal warping, are utilized to artificially expand limited datasets~\cite{dash2025silent}. % xuqisheng: 已加入表格
Transfer Learning and Domain Adaptation (e.g., Domain-Adversarial Training) enable the alignment of features across different subjects, facilitating the development of ``speaker-independent'' systems that require minimal recalibration. Furthermore, physiological normalization through Procrustes matching or feature-space Maximum Likelihood Linear Regression (fMLLR) helps reduce anatomical variance between individuals. A final significant advancement is the integration of Large Language Models (LLMs) as a post-processing stage; by rescoring the candidate output of the SSR decoder based on semantic probability, LLMs effectively correct phonetic hallucinations and significantly lower the final WER~\cite{benster2024cross}. % xuqisheng: 已加入表格

\subsection{Direct Silent Speech Synthesis: The Articulation-to-Acoustic}

The Articulation-to-Acoustic (ATA) paradigm, or Direct Synthesis, represents a sophisticated branch of SSI research designed to map physiological biosignals directly onto acoustic waveforms. Unlike recognition-based systems that rely on intermediate text decoding, ATA frameworks employ continuous mapping functions. This approach significantly minimizes processing latency and preserves paralinguistic nuances-such as idiosyncratic vocal identity and emotional prosody-that are frequently discarded in text-constrained pipelines~\cite{gosztolya2019autoencoder}. As evidenced by Table \ref{tab:ata_synthesis}, the field has transitioned from rigid statistical mappings to generative architectures capable of high-fidelity spectral reconstruction.

\newcolumntype{Y}[1]{>{\raggedright\arraybackslash\hsize=#1\hsize}X}

\begin{table*}[t]
\centering
\caption{Benchmarks of Direct Silent Speech Synthesis: Articulation-to-Acoustic (ATA) Paradigm}
\label{tab:ata_synthesis}
\footnotesize 
\renewcommand{\arraystretch}{1.2} 
\setlength{\tabcolsep}{4pt} % 略微增加列间距，利用全宽空间

% 总共有 4 个 Y 型列，权重之和：1.4 + 1.1 + 0.8 + 0.7 = 4.0
\begin{tabularx}{\textwidth}{@{} l Y{1.4} Y{1.1} c c c Y{0.9} Y{0.6} @{}}
\toprule
\textbf{Sensor Type} & \textbf{Model Architecture} & \textbf{Dataset / Corpus} & \textbf{Vocab.} & \textbf{Metric} & \textbf{Result Value} & \textbf{Population} & \textbf{Ref.} \\ \midrule

\textit{Neuro-Physio.} & & & & & & & \\
Neurotrophic Electrode & Kalman filter neural decoder & Vowel production & 3 Vowels & Hit Rate & 70.0\% & Locked-in & \cite{brumberg2010brain} \\
ECoG & Bi-LSTM (Seq2Seq) & Continuous Sent. & Open & MCD & 4.66--5.07 dB & Epilepsy & \cite{anumanchipalli2019speech} \\ \midrule

\textit{Kinematic} & & & & & & & \\
sEMG & DiffMV-ETS (Diffusion) & EMG-VCTK & Sentences & CER / nMOS & 45.8\% / 3.0 & Healthy & \cite{scheck2025diffmv} \\
sEMG & DNN / GMM conversion & Arctic / TIMIT & 108 Words & WER / MCD & 7.3\% / 4.5 dB & Healthy & \cite{janke2017emg} \\
sEMG & Tattoo-like patch + LDA & Instructions & 11 Cmds & Acc. & 89.0--92.3\% & Lar. & \cite{liu2020epidermal} \\
sEMG & Transformer + WaveNet & Sentences (Books) & Open & WER & 42.2\% & Healthy & \cite{gaddy2021improved} \\
PMA & RNN (GRU) / MFA & Arctic / TIDigits & 48 Items & Acc. / MCD & 91.5\% / 4.9 dB & Healthy & \cite{gonzalez2016silent, gonzalez2017direct} \\
\midrule

\textit{Imaging} & & & & & & & \\
UTI & Conformer + HiFi-GAN & Ultrasuite-Tal80 & 728 Utter. & MCD & 3.03--4.13 dB & Healthy & \cite{ibrahimov2025conformer} \\
UTI & CNN + WaveGlow & TaL / PPSD & Open & MCD & 5.37 dB & Healthy & \cite{csapo2020ultrasound} \\
UTI & DDPM (Diffusion)+PWG & TaL80 & Utterances & MCD / WER & 3.53 dB / 73.3\% & Healthy & \cite{zheng2024speech} \\
UTI + Video & Full-covariance HMM/GMM & French Sentences & Open & Acc. / MCD & 60\% / 5.2 dB & Healthy & \cite{hueber2016statistical} \\
UTI + Optical & MLP mapping (Eigentongue) & IEEE/Harvard & 720 Sent. & NMSE & 11.0\%--16.0\% & Healthy & \cite{hueber2007eigentongue} \\ \midrule

\textit{Acoustic / RF} & & & & & & & \\
Ultrasound & UNet + Transformer + PWG & LJSpeech / TIMIT & N/A & PESQ / MOS & 3.10 / 4.48 & Healthy & \cite{yu2025uspeech} \\
NAM / EL & GMM Voice Conversion (VC) & ATR Phoneme & 140 Sent. & MCD / MOS & 4.65 dB / 3.8 & Healthy/Lar. & \cite{hirahara2010silent} \\ 

\bottomrule
\end{tabularx}
\end{table*}

\subsubsection{Statistical Foundations and Early Neural Mapping}

For neuro-physiological signals, early intracortical systems established a baseline using linear state-space decoders, where Kalman filters mapped low-dimensional neural trajectories to acoustic parameters for locked-in patients \cite{brumberg2010brain}. As the paradigm expanded to kinematic and imaging modalities, researchers adapted statistical voice conversion frameworks-primarily GMM, and to estimate the joint probability density of articulatory-to-spectral mappings hirahara2010silent. To ensure acoustic continuity, generative HMM-based models were subsequently utilized to impose explicit temporal constraints, predicting smooth parameter paths like Mel-Cepstrum trajectories rather than independent frames \cite{hueber2016statistical}. Despite their foundational role, these statistical methods often struggled with the complex, non-linear dynamics of speech production. This limitation was particularly evident in high-dimensional UTI data, leading to the landmark Eigentongue decomposition, which applied PCA to project raw ultrasound images into a compact articulatory subspace \cite{hueber2007eigentongue}. This evolution toward compact representations motivated the adoption of DNN and ``bottleneck'' architectures \cite{janke2017emg}. By forcing networks to internalize informative latent representations of articulatory input, these multi-layer perceptrons significantly improved mapping accuracy and set the stage for modern generative synthesis.

% \textbf{GMM} were initially employed to estimate the joint probability density of articulatory-to-spectral mappings, a technique largely adapted from the field of voice conversion~\cite{csapo2017dnn,li2024end,yu2025silent}. To capture the essential temporal trajectories of speech, generative \textbf{HMM} were utilized to predict continuous parameter paths, such as Mel-Cepstrum or Line Spectral Pairs, rather than discrete phonetic states~\cite{hueber2016statistical,hueber2012continuous}. While these methods provided a foundational baseline, their linear nature struggled with the complex, non-linear dynamics of speech, leading to the adoption of Feed-Forward Neural Networks (FFNN) and ``bottleneck'' architectures~\cite{csapo2017dnn,xia2024systematic,gosztolya2019autoencoder,gosztolya2020applying}. These multi-layer perceptrons improved frame-by-frame mapping accuracy by forcing the network to internalize compact, informative latent representations of the articulatory input before expanding them into high-dimensional acoustic features.

\subsubsection{Sequence Modeling and Generative Adversarial Networks}
To resolve the challenges of co-articulation, where acoustic output is contingent upon surrounding articulatory states, sequence-aware models became the architectural standard. Bi-LSTM networks and Gated Recurrent Units (GRU) were successfully deployed to decode complex temporal dependencies in neuro-physiological and kinematic signals \cite{anumanchipalli2019speech, gonzalez2016silent}. Current high-performance backbones further incorporate Transformer and Conformer layers to leverage global self-attention for superior image and acoustic transduction \cite{ibrahimov2025conformer, yu2025uspeech}. To mitigate the ``over-smoothing'' effect inherent in traditional regression, Generative Adversarial Networks (GANs) are utilized within the synthesis pipeline. Specifically, neural vocoders such as HiFi-GAN and Parallel WaveGAN (PWG) are integrated to transform intermediate spectral features into high-fidelity audible waveforms \cite{ibrahimov2025conformer, zheng2024speech, yu2025uspeech}.

% To resolve the challenges of \textbf{co-articulation}, where acoustic output is contingent upon both preceding and subsequent articulatory states, sequence-aware models became the architectural standard~\cite{gosztolya2019autoencoder}. \textbf{Bidirectional Long Short-Term Memory (Bi-LSTM)} networks process articulatory sequences in both temporal directions to produce smoother acoustic trajectories than frame-based approaches~\cite{cao2018articulation,liu2018articulatory,kim2017speaker}. For image-centric modalities such as UTI, \textbf{Convolutional Recurrent Neural Networks (CRNN)} are employed to synthesize spatial feature extraction via CNNs with the temporal modeling of LSTMs~\cite{xia2024systematic,shandiz2021improving}. A persistent hurdle in regression-based ATA is the ``over-smoothing'' effect inherent in Mean Squared Error (MSE) loss, which often results in muffled, ``buzzy'' audio. To mitigate this, \textbf{Generative Adversarial Networks (GANs)} have been successfully integrated into systems like \textit{Vid2Speech}. By utilizing a discriminator to penalize outputs that deviate from the distribution of natural speech spectrograms, GANs encourage the generator to produce high-frequency spectral details that traditional regression models miss~\cite{reed2016generative}.

\subsubsection{Diffusion Models and Neural Vocoding}
The current technical frontier in ATA synthesis has shifted toward Denoising Diffusion Probabilistic Models (DDPM) \cite{zheng2024speech}. Architectures such as \textit{DiffMV-ETS} learn to reverse a gradual Gaussian noising process, iteratively refining a signal from pure noise into a high-fidelity \textit{mel}-spectrogram conditioned (directly or indirectly) on articulatory input \cite{scheck2025diffmv}. Diffusion models are increasingly favored due to their superior ability to generate less over-smoothed and more diverse acoustic features than previous non-probabilistic mapping models \cite{zheng2024speech, scheck2025diffmv}. Crucially, the diffusion model first predicts an intermediate spectral representation, which is then transformed into an audible waveform by a vocoder \cite{zheng2024speech}. While earlier pipelines often relied on parametric/statistical vocoding (e.g., predicting MFCC/$F_0$ and using MLSA), which requires explicit excitation-related features such as fundamental frequency ($F_0$) \cite{janke2017emg}, modern frameworks increasingly employ neural vocoders such as WaveNet \cite{gaddy2021improved}, WaveGlow \cite{csapo2020ultrasound}, or HiFi-GAN \cite{ibrahimov2025conformer}. These neural decoders fill in the fine-grained details required for high intelligibility and naturalness.

\subsubsection{Training Algorithmic Innovations}
A critical challenge in training ATA models for `silent' speech is the inherent lack of ground-truth audio targets. To address this, researchers have pioneered Pseudo-Target Generation. The system is initially trained on vocalized data where both articulatory and acoustic signals coexist; subsequent silent articulatory data is then paired with `hallucinated' audio generated via alignment algorithms like DTW or synthesized through Text-to-Speech (TTS) engines. This synthetic supervision allows the ATA model to refine its mapping capabilities even in the absence of original vocalized recordings, facilitating the transition from laboratory prototypes to practical clinical applications~\cite{zheng2024speech,scheck2025diffmv}.

\subsection{Linguistic Priors and the LLM Frontier}

Historically, the computational bottleneck of SSIs has resided in the inherent ambiguity and non-stationarity of biosignals. The recent adoption of LLMs as the semantic engine of SSI systems has catalyzed a paradigm shift, as summarized in Table III. Rather than relying solely on acoustic or articulatory modeling, modern frameworks leverage the vast linguistic priors of LLMs to perform latent semantic alignment, transforming fragmented physiological gestures into coherent communication.

\begin{table*}[!htbp] 
\centering 
\caption{Linguistic Priors and the LLM Frontier: Categorized Benchmarks.}
\begin{tabularx}{\textwidth}{XXXXXXXl}
\toprule
\textbf{Sensor Type} & \textbf{Model Architecture} & \textbf{Dataset / Corpus} & \textbf{Vocab.} & \textbf{Metric} & \textbf{Result Value} & \textbf{Population} & \textbf{Ref.} \\ \toprule

\multicolumn{8}{l}{\textit{Neuro-Physio.}} \\ \midrule
MEA & RNN + n-gram LM + LLM rescoring & Prompted sentences & 125k words & WER & 2.5\% & ALS Patient & \cite{kimura2022silentspeller} \\
EEG & Transformer + ResBlocks + CTC & Alice Audiobook & 601 words & Top-10 Acc. & 26.82\% & Healthy & \cite{chen2025decoding} \\ \midrule

\multicolumn{8}{l}{\textit{Kinematic}} \\ \midrule
sEMG + Audio & MONA LISA (GPT-3.5 fine-tuned) & Gaddy benchmark & Open & WER & 12.2\% & Impaired & \cite{benster2024cross} \\
sEMG & Transformer + GPT-2 post-proc. & Digital Voicing (Gaddy) & - & WER & 30\% & - & \cite{sivasubramaniam2025silent} \\
sEMG & HMM-DNN + Bundled Phonetic Feat. & - & - & WER & 34.7\% (Silent) & Healthy & \cite{gonzalez2020silent} \\
Facial sEMG & Transformer + CTC + Aux. Tasks & NBA EMG-to-Text & 667 char. & CER & 38.0\% & Healthy & \cite{xie2025neural} \\
sEMG & EMG Adaptor + Llama3-3B & Gaddy \& Klein (2021) & 67 words & WER & 0.49 & Healthy & \cite{mohapatra2025can} \\
EMG + PZT & Siamese Network (SNN) + PMLDF & Alphabet letters & 8 letters & Acc. & 95.63\% & Healthy & \cite{kang2025wearable} \\ \midrule

\multicolumn{8}{l}{\textit{Imaging}} \\ \midrule
Video (Lip) & LipType (LLaMA-3.2-3B) & LRS3 test set & Large & WER & 9.19\% & Healthy & \cite{su2025multimodal} \\
Video (Lip) & Auto-AVSR & LRS3 test set & Large & WER & 20.3\% & Healthy & \cite{su2025multimodal} \\
Ultrasound & ViT-based Self-supervised & Eshky et al. (2018) & - & Acc. & 88.94\% & Children & \cite{xia2024systematic} \\
Ultrasound & CNN-based Self-supervised & - & - & Acc. & 79.4\% & Healthy & \cite{xia2024systematic} \\ \midrule

\multicolumn{8}{l}{\textit{Acoustic / RF}} \\ \midrule
Ultrasound + Video & HMM-based + HNM & CMU ARCTIC & 3k words & Acc. & 83.3\% & Healthy & \cite{hueber2011statistical} \\ \bottomrule

\end{tabularx}
\end{table*}

\subsubsection{Functional Taxonomy: The Evolving Role of LLMs}
The functional integration of LLMs within the SSI pipeline has evolved from peripheral post-processing to core signal decoding. Initially, LLMs served as high-level ``semantic revisors.'' For instance, \cite{sivasubramaniam2025silent,benster2024cross} introduced a dual-stage framework where a Transformer-based ASR converts signals to noisy text, followed by an LLM that resolves syntactic inconsistencies and phonetic ambiguities. A more sophisticated approach, exemplified by the MONA LISA system \cite{benster2024cross}, utilizes LLM Integrated Scoring Adjustment (LISA) to rescore the $N$-best candidates from a beam search, prioritizing semantic plausibility over raw signal likelihood. Beyond correction, recent research explores Direct Signal-to-Semantic Translation, where physiological features are treated as ``Neural Tokens.'' \cite{mohapatra2025can} demonstrated an EMG Adaptor that maps sEMG features directly into the embedding space of a Llama-3 backbone, allowing the model to interpret muscle activity as a high-dimensional dialect. This transition enables Zero/Few-shot Personalization, addressing the chronic scarcity of individualized data. The \textbf{LipLearner} system \cite{su2023liplearner} achieves personalized command recognition with as little as six minutes of training data by exploiting linguistic transfer learning. Furthermore, natural language prompts are now utilized to condition the LLM on specific modalities; for example, specific prefix prompts such as \textit{$Unvoiced EMG: <emg_embed> Prompt: Convert to text$} are employed to guide the model’s attention toward articulatory-specific latent representations \cite{mohapatra2025can}.

\subsubsection{Architectural Convergence: Alignment and Fusion Mechanisms}
The primary technical challenge remains the modality alignment problem-projecting continuous, high-frequency bio-signals into discrete semantic latent spaces. Current state-of-the-art (SOTA) systems utilize 1D-CNN or ResNet-based adaptors to downsample high-frequency signals, such as sEMG or EEG, and project them into the LLM’s embedding dimension. To bridge the ``Silent Gap,cross-contrastive loss ($crossCon$) and supervised temporal contrastive loss ($supTcon$) align silent signal manifolds with their voiced counterparts in a shared latent space. Architectural trends show a convergence toward hybrid models, typically pairing a Conformer visual/signal encoder with a Llama-3.2 (1B or 3B) decoder to balance performance with mobile-edge deployment requirements. To maintain the LLM's vast knowledge while adapting to niche physiological data, Parameter-Efficient Fine-Tuning (PEFT), specifically Low-Rank Adaptation (LoRA), has become the standard, achieving high SSI accuracy without the ``catastrophic forgetting'' of underlying linguistic foundations~\cite{su2025multimodal}.

\subsubsection{Performance Benchmarks and the Usability Threshold}
The integration of LLMs has led to a historic reduction in WER, particularly in open-vocabulary scenarios. A significant milestone was achieved by the MONA LISA system, which reduced the WER on the Gaddy (2020) benchmark from 28.8\% to 12.2\%. This represents the first time a non-invasive sEMG-based SSI has broken the 15\% ``usability threshold'' required for real-world application. This gain in accuracy, however, comes with a significant Latency Cost. While models like Llama-3.2-1B are optimized for edge devices, using massive LLMs (e.g., GPT-4) for rescoring introduces substantial inference delays, highlighting a critical trade-off between semantic depth and real-time interaction.

\subsubsection{The Semantic Chasm and Future Frontiers}

Despite these gains, the ``Linguistic Frontier'' faces two critical challenges: the Data Chasm and the Hallucination Bottleneck. The former refers to the fundamental mismatch between trillion-token text corpora and hour-scale physiological datasets. Current methods rely heavily on parallel ``voiced'' data as a bridge, which is often unavailable for aphonic patients. The latter refers to the risk of Over-correction; if the input signal is too degraded, the LLM may ``hallucinate'' a grammatically perfect sentence that completely deviates from the user's actual intent \cite{su2025systematic}. Recently, the trend is moving toward Large Brain Language Models (LBLMs) \cite{zhou2025mindspeak}. These represent a shift from ``adapting LLMs'' to ``building native physiological LLMs,'' where models are pre-trained directly on massive, self-supervised bio-signal datasets. This approach aims to resolve the alignment problem by treating human neural activity not as a noise-ridden input to be translated, but as a primary, native linguistic modality.

Looking forward, the trend is moving toward Large Brain Language Models (LBLMs). These represent a shift from ``adapting LLMs'' to building ``native physiological LLMs'' pre-trained directly on massive, self-supervised bio-signal datasets. This approach aims to resolve the alignment problem by treating human neural activity not as a noise-ridden input to be translated, but as a primary, native linguistic modality. By shifting the burden of reconstruction from noisy physical sensors to robust linguistic priors, the field has finally approached the 15\% WER threshold, moving SSIs from laboratory curiosities to viable assistive technologies.

%%%%%%%%%%%%%%%%%%%%%%%%%%%%%%%%%%%%%%%%%%%%%%%%%%%%%%%%%%%%%%%%%%%%%%%%%%%
% SECTION IV: BENCHMARKING - METRICS AND DATASETS
%%%%%%%%%%%%%%%%%%%%%%%%%%%%%%%%%%%%%%%%%%%%%%%%%%%%%%%%%%%%%%%%%%%%%%%%%%%

\section{Benchmarking: Evaluation Metrics and Open Datasets}
\label{sec:benchmarking}
The progression of SSIs toward reproducible development relies on a standardized evaluative architecture. This section establishes a framework for assessing SSIs across recognition and synthesis modalities, bridging the gap between raw signal acquisition and communicative utility.

\subsection{Quantitative Performance Metrics}

Evaluating an SSI requires a multi-dimensional lens that integrates linguistic precision, acoustic naturalness, and operational efficiency. Because these interfaces map non-acoustic biosignals-such as sEMG, Ultrasound, or EEG-onto high-level semantic units, simplistic accuracy metrics often fail to capture the nuances of the human-in-the-loop experience \cite{abdullah2020computationally}. Consequently, researchers employ a synthesis of information theory and psychoacoustic models to gauge restorative efficacy for clinical cohorts \cite{benster2024cross, tang2024ultrasensitive}.

\subsubsection{Linguistic Accuracy (Recognition Metrics)} 

Word Error Rate (WER) remains the primary benchmark for recognition-based decoders, derived from the Levenshtein distance:\begin{equation}\text{WER} = \frac{S + D + I}{N} \times 100\end{equation}where $S, D, I$ denote substitutions, deletions, and insertions, and $N$ is the reference count \cite{zeng2023msilent}. A WER below the 15\% threshold is generally deemed essential for SSIs to achieve functional parity with traditional ASR \cite{benster2024cross}. Given the phonetic density of silent articulation, research increasingly adopts Phoneme (PER) or Character Error Rate (CER) to resolve fine-grained ambiguities, especially for out-of-vocabulary (OOV) terms and tonal languages like Mandarin \cite{hahm2015silent, wang2024watch}. Beyond error rates, communicative throughput is quantified by the Information Transfer Rate (Wolpaw Rate), with state-of-the-art neural decoders now approaching natural conversational speeds of 100 WPM \cite{wadkins2019continuous}.

\subsubsection{Acoustic and Perceptual Fidelity (Synthesis Metrics)}

Direct synthesis (ATA) systems utilize a hybrid of objective signal processing and subjective perception to evaluate reconstructed waveforms. Unlike recognition, ATA seeks to preserve prosodic cues by modeling nonlinear articulatory-to-acoustic manifolds \cite{gosztolya2019autoencoder}. Specifically, Mel-Cepstral Distortion (MCD) serves as the standard objective measure for spectral similarity:\begin{equation}MCD[dB]=\frac{10}{\ln 10} \sqrt{2 \sum_{d=1}^{D} (c_d - \hat{c}_d)^2}\end{equation}where $D$ denotes the feature dimensionality. While high-fidelity systems targeting UTI or EMA signals report MCD values between 4.5 dB and 6.5 dB, objective gains do not always align with intelligibility due to the ``averaging effect'' of statistical models \cite{hueber2016statistical}.To better approximate human audition, PESQ and STOI are employed to measure naturalness and temporal envelope correlation, respectively \cite{zhou2025m2silent}. Modern frameworks further utilize Extended STOI (ESTOI) to maintain phonetic contrasts lost in spectral regression. Furthermore, recovering Prosodic Fidelity remains a major challenge due to the lack of a direct anatomical link between vocal fold vibration and supraglottal motion \cite{xia2024systematic}. Current systems utilize auxiliary modules to predict $F_0$ trajectories and voiced/unvoiced (V/UV) states to avoid robotic monotony \cite{grosz2018f0}. Finally, the Mean Opinion Score (MOS) remains the gold standard for subjective validation \cite{zheng2023speech}, with a recent shift toward the MUSHRA protocol to achieve higher statistical sensitivity against hidden anchors \cite{gosztolya2019autoencoder}.

\subsubsection{System-Level Constraints}Practical deployment necessitates benchmarking against Inference Latency and the Real-Time Factor (RTF). For seamless interaction, end-to-end delays must remain below 50 ms to prevent speech disfluency \cite{hernaez2021voice}. Wearable research further prioritizes Power Consumption (mW) and Computational Complexity (FLOPs), with low-power platforms now maintaining acquisition within a 22.2 mW to 73.3 mW envelope \cite{meier2025parallel, zhang2023echospeech}.

\subsection{Open-Source Datasets and Repositories}

The shift from closed-set laboratory experiments to large-scale, open-vocabulary decoding has been driven by the release of several pivotal multimodal repositories \cite{kimura2022ssr7000}. These datasets address core SSI challenges, such as the ``Silent Lombard Effect'' and the articulatory mismatch between vocalized and silent modes \cite{beeson2023silent}. Among sEMG resources, the Gaddy dataset (19 hours) remains a cornerstone for digital voicing through parallel silent/voiced recordings \cite{gaddy2021improved}. In the imaging domain, the Silent Speech Challenge  archive provides over 700,000 frames of synchronized ultrasound and lip images \cite{wang2021representation}, while LRS3-TED serves as the standard ``in-the-wild'' benchmark for visual speech recognition with over 400 hours of data \cite{yang2025audio}. Recent advancements have introduced large-scale Mandarin and neural-based datasets. The AVE Speech dataset offers 55 hours of synchronized audio, video, and 6-channel EMG from 100 participants, which is critical for studying tonal languages \cite{zhou2025ave}. In the neural frontier, the MindSpeak 2025 repository (120 hours of HD-EEG) facilitates the self-supervised pre-training of Large Brain Language Models (LBLM) \cite{zhou2025pretraining}. Parallel to these, EchoSpeech leverages active acoustic sensing on eyewear, demonstrating high-accuracy articulatory tracking at a low power profile of 73.3 mW \cite{zhang2023echospeech}. Specialized repositories like SSR7000 focus on ``out-of-the-lab'' variability, providing raw data to evaluate model robustness against sensor misalignment \cite{kimura2022ssr7000}. Clinical restorative applications are supported by the MultiNAM dataset and the ReSSInt project, which target laryngectomy and ALS rehabilitation through high-fidelity neural vocoders that preserve speaker identity and expressive prosody \cite{shah2025advancing, hernaez2021voice}.

\begin{table*}[t]
\centering
\caption{Summary of Representative Open-Source Datasets for Silent Speech Research.}
\label{tab:datasets}
\begingroup
\renewcommand{\arraystretch}{1.1}
\scriptsize
\setlength{\tabcolsep}{3pt}
\sloppy
\hyphenpenalty=10000
\exhyphenpenalty=10000
\begin{tabularx}{\textwidth}{@{}
p{2cm}
p{1.8cm}
p{0.9cm}
p{1.8cm}
p{1.6cm}
>{\raggedright\arraybackslash}p{5.8cm}
>{\raggedright\arraybackslash}p{2.8cm}
@{}}
\toprule
\textbf{Dataset} & \textbf{Modality} & \textbf{Subj.} & \textbf{Vocabulary} & \textbf{Duration} & \textbf{Key Features} & \textbf{Dataset Source / Link} \\
\midrule
\multicolumn{7}{@{}l@{}}{\textit{I. Surface Electromyography (sEMG) Datasets}} \\
\addlinespace[4pt]
EMG-UKA (Full / Trial) \cite{wand2014emg} & sEMG + Audio & 8 / 4 & 2,102 / 108 Words & 7.7 / 1.87 Hours & Multi-mode (audible, whispered, silent); Broadcast News domain. Public subset for feature extraction benchmarking. & \url{csl.anthropomatik.kit.edu} \\
Gaddy sEMG  \cite{gaddy2020digital,gaddy2021improved} & sEMG + Audio & 1 & 67 Words & 18.6 Hours & Breakthrough in digital voicing with paired voiced/silent recordings. & \url{zenodo.org/records/4064409} \\
AVE Speech \cite{zhou2025ave} & A + V + EMG & 100 & 100 Sent. & 55.3 Hours & First large-scale Mandarin sentence-level EMG-visual corpus. & \url{huggingface.co/datasets/MML-Group/AVE-Speech} \\
EMG2QWERTY \cite{sivakumar2024emg2qwerty} & Wrist sEMG & 10+ & QWERTY Keys & Large Scale & Touch typing decoding task using wearable sEMG wristbands. & \url{huggingface.co/datasets/cyrilzakka/emg2qwerty} \\
\midrule
\multicolumn{7}{@{}l@{}}{\textit{II. Ultrasound Tongue Imaging (UTI) and Visual (VSR) Datasets}} \\
\addlinespace[4pt]
SSC Archive \cite{denby2004speech,cai2011recognition} & UTI + Optical & 1 & 5,000 Words & 700k Images & A landmark benchmark providing synchronized tongue ultrasound and lip video based on TIMIT and WSJ0 corpora. & \url{ftp.espci.fr/pub/sigma/} \\
SSR7000 \cite{kimura2022ssr7000} & UTI + Optical & 1 & Open Vocab & 7,484 Utter. & Large-scale synchronized corpus optimized for E2E SSR; includes raw, pre-processed, and feature data. & \url{github.com/supernaiter/ssr7000} \\
TaL \cite{ribeiro2021tal} & UTI + V + A & 82 & Open Vocab & 24 Hours & Multi-speaker synchronized corpus featuring one professional (TaL1) and 81 regular speakers (TaL80) in modal and silent modes. & \url{ultrasuite.github.io/data/tal_corpus/} \\
UltraSuite \cite{eshky2019ultrasuite} & UTI + Audio & 113 & Mixed & 18.7 Hours & Clinical database focused on child speech therapy; includes data from typical and disordered speech development. & \url{ultrasuite.github.io} \\
AUSpeech \cite{yang2025audio} & UTI + A + Text & 54 & Mandarin Sents. & 22.31 Hours & First high-res (920$\times$700) Mandarin audio-ultrasound sync database covering all permissible syllable onsets and daily sentences. & \url{cstr.cn/31253.11.sciencedb.18722} \\
LRS3-TED \cite{afouras2018lrs3} & Visual (VSR) & Thousands & Open Vocab & 400+ Hours & Largest ``in-the-wild'' visual speech dataset; core resource for large-scale lip-reading pre-training and LLM-enhanced models. & \url{robots.ox.ac.uk/~vgg/data/lrs3/} \\
LRW-1000 \cite{yang2019lrw} & Visual (VSR) & 2000 & 500-1k Words & 1M+ Instances & Large-scale naturally-distributed benchmarks for word-level English (LRW) and Mandarin (LRW-1000) visual speech recognition. & \url{vipl.ict.ac.cn/view_database.php} \\
\midrule
\multicolumn{7}{@{}l@{}}{\textit{III. EEG and Neural Signal Datasets}} \\
\addlinespace[4pt]
Thinking Out Loud \cite{nieto2022thinking} & Neural (EEG) & 10 & 5 Words & Multi-session & Open-access BCI dataset for Spanish inner speech recognition. & \url{github.com/N-Nieto/Inner_Speech_Dataset} \\
Alice EEG \cite{brennan2019hierarchical} & Neural (EEG) & 33 & Open (Story) & 12.4 Mins & Passive audiobook listening for linguistic prediction analysis. & \url{deepblue.lib.umich.edu/data/concern/data_sets/bg257f92t} \\
SparrKULee \cite{accou2024sparrkulee} & Auditory EEG & 85 & Continuous & 168 Hours & Large auditory-evoked response repository for speech stimuli. & \url{rdr.kuleuven.be/dataset.xhtml?persistentId=doi:10.48804/K3VSND} \\
\midrule
\multicolumn{7}{@{}l@{}}{\textit{IV. Magnetic Sensing and Articulatory Imaging (EMA/PMA/MRI)}} \\
\addlinespace[4pt]
MOCHA-TIMIT \cite{wrench2000multichannel} & EMA + EPG + A & 2 & 460 Sent. & 920 Utter. & Classic articulatory database used for coarticulation modeling. & \url{cstr.ed.ac.uk/research/projects/artic/mocha.html} \\
USC-TIMIT \cite{narayanan2014real} & rtMRI + Audio / EMA + Audio & 10 / 4 & 460 Sent. & 3.05 Hours / 460 Utter. & High-quality vocal tract scans with paired denoised audio; Fleshpoint tracking data parallel to the MRI corpus. & \url{sail.usc.edu/span/usc-timit/} \\
MNGU0 \cite{richmond2011announcing} & EMA + Audio & 1 & Open Vocab & 67 Minutes & Single speaker EMA data with Day 1/Day 2 subsets. & \url{mngu0.org} \\
\midrule
\multicolumn{7}{@{}l@{}}{\textit{V. Acoustic, Radar, and Other Sensing Modalities}} \\
\addlinespace[4pt]
MultiNAM \cite{shah2025advancing} & NAM + V + A & 2 & 21 Utter. & 7.96 Hours & Paired non-audible murmur, facial video, and whispers. & \url{diff-nam.github.io/DiffNAM/} \\
TAPS \cite{kim2025taps} & Throat + Audio & 60 & 4,000 Utter. & 15.3 Hours & Paired supraglottic throat mic and acoustic recordings. & \url{huggingface.co/datasets/yskim3271/Throat_and_Acoustic_Pairing_Speech_Dataset} \\
\bottomrule
\end{tabularx}
\endgroup
\end{table*}

\subsection{Standardization and Generalization Hurdles}

Despite the expansion of open-source repositories, achieving robust standardization in SSIs remains obstructed by several fundamental bottlenecks.

First, Anatomical Heterogeneity across clinical and healthy cohorts-encompassing variance in lingual morphology and neuromuscular composition-induces significant distribution shifts in biosignals. This inter-subject variability renders zero-shot, subject-independent transfer a persistent challenge, necessitating complex domain adaptation strategies to align disparate user manifolds \cite{gowda2025non, xia2024systematic}. 
Second, the system must navigate Cross-modal and Session Mismatch. The absence of auditory feedback in silent modes often triggers the ``Silent Lombard Effect,'' characterized by reduced articulatory effort and hypo-articulation. Furthermore, the high sensitivity of wearable sensors to sub-millimeter repositioning between sessions causes severe signal misalignment, complicating the longitudinal stability of decoding models \cite{stone2021silent, ribeiro2021silent}. 
Finally, a critical Annotation and Alignment Gap persists due to the lack of universal protocols for synchronizing high-dimensional imaging with discrete semantic units. Mapping continuous ultrasound sequences to acoustic or linguistic features requires sub-millisecond precision, yet the informational asymmetry between articulatory dynamics and target phonemes often leads to temporal jitter and decoding artifacts \cite{ramanarayanan2018analysis, yu2025uspeech}.

\section{Applications of SSIs}
\label{sec:applications}

The maturation of SSIs has extended their utility beyond acoustic-free recognition. By decoupling linguistic expression from vocal fold vibration, SSIs facilitate a fundamental shift in how information is externalized, categorized here through clinical restoration, clandestine interaction, and ubiquitous computing.

\begin{figure*}[t]
\centering
\includegraphics[width=1.0\linewidth]{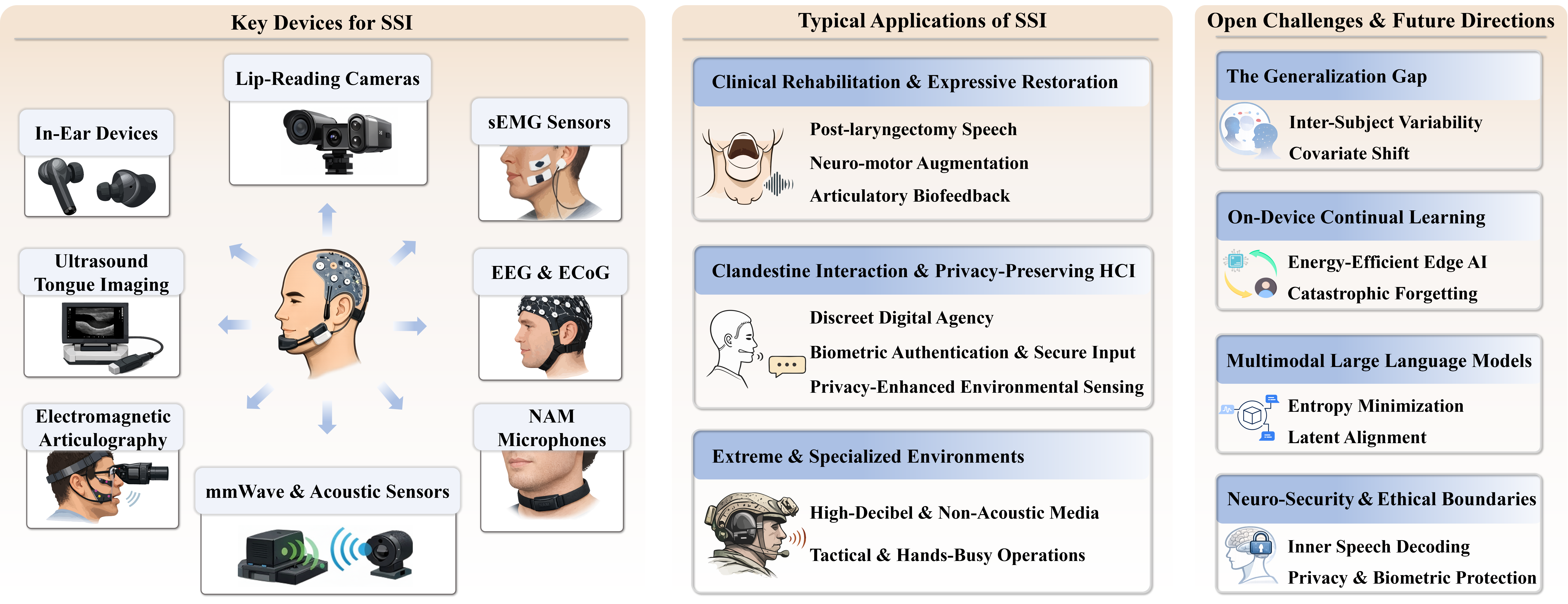} % 请确保路径正确
\caption{Overview of Silent Speech Interface (SSI): Key sensing modalities, typical application domains, open challenges and future directions. This taxonomy bridges the gap between raw physiological signal interception and high-bandwidth communicative utility across clinical and ubiquitous computing contexts.}
\label{fig:nma_chain}
\end{figure*}

\subsection{Clinical Rehabilitation and Expressive Restoration}The primary imperative for SSI research is restoring communicative agency for individuals with severe phonation disorders.

\subsubsection{Post-Laryngectomy Vocal Rehabilitation}Total laryngectomy deprives patients of their primary sound source, often leading to social isolation. Traditional alaryngeal methods, such as the electrolarynx, are frequently criticized for their robotic quality and low social acceptability \cite{cao2023magtrack}. SSIs offer a superior alternative by mapping intact articulatory control directly to personalized speech synthesis \cite{gonzalez2020silent}. Modern systems utilize voice banking to drive neural vocoders, restoring the user's idiosyncratic timbre and vocal identity with latencies under 50 ms-essential for maintaining the speaker's auditory feedback loop \cite{juanpere2019ultrasound, meltzner2017silent}.

\subsubsection{Neuro-motor Augmentation}

For individuals with ALS or Locked-in Syndrome, traditional Augmentative and Alternative Communication (AAC) devices are restricted by low throughput ($<10$ WPM) and high cognitive fatigue \cite{stavisky2025restoring}. SSIs enable high-bandwidth communication by decoding subvocal articulatory intent. High-resolution intracortical BCIs have demonstrated conversational throughput exceeding 60 WPM with $>99\%$ accuracy \cite{finkelstein2025breaking}. Advanced ECoG-based decoders provide stable, long-term performance, allowing paralyzed users to independently manage personal computers and maintain professional employment.

\subsubsection{Articulatory Biofeedback}

In speech therapy and second-language (L2) acquisition, SSIs provide real-time visual-acoustic biofeedback. Disorders like apraxia involve difficulty coordinating invisible articulators. By visualizing lingual dynamics via ultrasound (UTI) or magnetic tracking (EMA), users can instantly observe discrepancies between their current gesture and the phonetic target \cite{xia2024systematic}. This recalibrates kinematic trajectories, improving intelligibility for stroke or oral cancer survivors \cite{garcia2024enabling, dai2021effects}.

\subsection{Clandestine Interaction and Privacy-Preserving HCI}

SSIs mitigate the ``broadcast nature'' of acoustic communication, enabling high-bandwidth interaction in noise-sensitive or high-security settings where traditional Voice User Interfaces (VUIs) are impractical.

\subsubsection{Discreet Digital Agency}

SSIs facilitate a ``human-machine cognitive coalescence'' by allowing users to interact with AI assistants via internal articulation-subtle activation of articulators without vocalization \cite{kapur2018alterego}. Systems such as AlterEgo capture residual neuromuscular signals, enabling bidirectional conversation in public spaces without alerting bystanders. This paradigm treats computing as a natural extension of the user's cognition, addressing the social awkwardness often associated with verbally addressing devices in public \cite{su2023liplearner, pandey2021acceptability}.

\subsubsection{Biometric Authentication and Secure Input}

Silent speech integrated with unique anatomical features enables robust security frameworks. Systems like \textbf{HEar-ID} capture Ear Canal Dynamic Motion (ECDM) as a biometric signature \cite{dong2025recognizing}, facilitating ``silent passwords'' resistant to acoustic eavesdropping or replay attacks. Furthermore, technologies like \textbf{ToothSonic} and \textbf{Lipwatch} utilize articulatory biometric signatures to verify identity during input, ensuring device access is granted only to legitimate users \cite{zhang2024lipwatch, dong2025recognizing}.

\subsubsection{Privacy-Enhanced Environmental Sensing}

Contactless SSIs using mmWave radar (\textit{mSilent}) or active acoustic sensing (\textit{EchoSpeech}) offer privacy-preserving alternatives to camera-based lip-reading \cite{zeng2023msilent}. Unlike RGB video, which captures identifiable facial textures, wireless sensing captures coarse-grained articulatory dynamics, minimizing biometric leakage while operating through obstructions such as face masks \cite{zhang2023echospeech, lee2025ir}.

\subsection{Environmental Resilience and Specialized Scenarios}

SSIs provide a solution for environments where traditional microphones are functionally insufficient by capturing the physiological \textit{causes} of speech rather than its acoustic \textit{consequences}.\subsubsection{High-Decibel and Non-Acoustic Media}Standard ASR degrades precipitously in extreme noise (SNR $<0$). SSIs remain immune by utilizing mmWave, sEMG, or Non-Audible Murmur (NAM) sensors to capture body-conducted vibrations \cite{shah2025advancing}. Modern skin-attached acoustic sensors (SAAS) maintain signal acquisition in 125 dB environments where air-conduction microphones fail \cite{liu2025machine}. Furthermore, SSIs enable verbal interaction in environments lacking an acoustic medium, such as underwater or extravehicular space operations \cite{dong2023decoding, xu2023force}.

\subsubsection{Tactical and Hands-Busy Operations}

For personnel in critical tasks (e.g., firefighters or soldiers), ``earable'' SSIs or textile-based sensors allow hands-free control without clear acoustic pathways. This is vital for users wearing breathing apparatuses or hazmat suits that cause significant acoustic obstructions \cite{jin2022earcommand}. Recent advancements enable the integration of multi-channel textile EMG sensors directly into earmuffs, achieving high accuracy for voice-free control while maintaining discretion \cite{tang2025wireless, dash2025silent}. In summary, SSIs function as a ``digital proxy'' for the vocal tract, capturing whispers of intent legible to the machine even when vocalization is restricted by physiology or context.

%%%%%%%%%%%%%%%%%%%%%%%%%%%%%%%%%%%%%%%%%%%%%%%%%%%%%%%%%%%%%%%%%%%%%%%%%%%
% SECTION V: OPEN CHALLENGES AND STRATEGIC ROADMAP
%%%%%%%%%%%%%%%%%%%%%%%%%%%%%%%%%%%%%%%%%%%%%%%%%%%%%%%%%%%%%%%%%%%%%%%%%

\section{Open Challenges and Future Directions}\label{sec:challenges}Despite the trajectory from laboratory prototypes to commodity wearables, the transition of SSIs into ubiquitous deployment remains attenuated by fundamental bottlenecks. Current research must transcend controlled environments to address the stochastic nature of human physiology and the rigorous constraints of edge computing. We delineate a strategic roadmap focusing on four orthogonal research frontiers.

\subsection{The Generalization Gap: Zero-Shot Transfer via Foundation Models}

The profound inter-subject variability inherent in physiological signals remains a persistent obstacle. Anatomical differences in lingual morphology and neuromuscular manifolds imply a severe Covariate Shift, where the input distribution $P(X)$ varies drastically across individuals even when the semantic intent $P(Y|X)$ is constant. To mitigate this, the paradigm is shifting toward Cross-modal Foundation Models. By pre-training on massive, unlabeled bio-signal corpora and acoustic data, these models can learn a universal articulatory latent space. Analogous to the impact of LLMs on natural language processing, we envision a ``plug-and-play'' experience enabled by Neuro-semantic Alignment, where a frozen pre-trained encoder maps a new user's biosignals into a shared embedding space. This zero-shot capability, supplemented by Meta-Learning (e.g., MAML) and Domain Adversarial Training, targets the extraction of features robust to electrode displacement and physiological non-stationarity, potentially eliminating user-specific calibration.

\subsection{On-Device Continual Learning and Hardware-Algorithm Co-Design}

To achieve natural prosody, SSIs must operate within a closed-loop latency of $\le 50$ ms. This necessitates a shift from cloud-based inference to hardware-aware Edge AI. However, static models fail to account for the temporal drift of biosignals over weeks of use. Future research must prioritize On-device Continual Learning (OCL)~\cite{taylorbefore,lyu2025mitigating}, enabling the interface to evolve locally without the risk of Catastrophic Forgetting.Central to this transition is the integration of energy-efficient Neuromorphic Computing. Spiking Neural Networks (SNNs) implemented on specialized hardware offer a trajectory for processing signals in an event-driven manner that mirrors biological neural activity~\cite{bouvier2019spiking}. By combining OCL with algorithmic strategies like Knowledge Distillation and Dynamic Pruning, Transformer-based decoders can be compressed into wearable form factors while maintaining the ability to fine-tune to the user's shifting physiology in real-time.

\subsection{Multimodal Large Language Models and Latent Synergy}

Single-modality SSIs are inherently constrained by the physical limitations of their interception points-such as sEMG’s sensitivity to impedance or ultrasound's susceptibility to acoustic shadows. The mathematical imperative for Multimodal Fusion is the minimization of posterior entropy: $H(Y|X_{\text{combined}}) \le \min(H(Y|X_{i}))$. A significant breakthrough is the transition toward Multimodal Large Language Models (MLLMs) for SSIs.By aligning continuous bio-signal frames (e.g., sEMG, UTI, EEG) with the embedding space of a pre-trained LLM via contrastive objectives, the system can leverage the model’s vast Semantic Prior to resolve phonetic ambiguities. Furthermore, addressing the ``Silent Lombard Effect''~\cite{luo2018lombard}-the articulatory drift caused by a lack of auditory feedback-requires closed-loop biofeedback. Future MLLMs should incorporate haptic or visual cues to assist users in maintaining stable articulatory targets within a unified latent manifold over prolonged use~\cite{huang2025lend}.

\subsection{Neuro-Security and Ethical Boundaries of Cognitive Liberty}

As SSIs advance toward decoding ``inner speech,'' they breach the critical boundary of Cognitive Liberty. Physiological signals contain latent identifiers, including emotional states and health pathologies, raising the stakes for data sovereignty. We introduce the concept of ``Neuro-security'' to define the technical safeguards required for biological data.Technical defenses must integrate Federated Learning (FL) and Differential Privacy (DP) to ensure that raw signals remain localized. Moreover, the vulnerability of SSI models to adversarial perturbations-where malicious actors might inject ``silent commands"-demands the development of robust, certified defense layers. To date, established ethical frameworks must prevent non-consensual ``neuro-surveillance,'' integrating hardware-level kill-switches to ensure that the decoding of private thoughts remains exclusively under volitional control.

\section{Conclusion} \label{sec:conclusion}

SSIs have reached a critical inflection point, transitioning from laboratory prototypes to high-performance, clinically viable benchmarks. To now, the field is defined by the convergence of three pillars: flexible bio-electronics for high-fidelity sensing, cross-modal latent alignment via Transformer-based architectures, and generative neural vocoders for expressive, low-latency synthesis. This metamorphosis from niche restorative tools to cornerstones of pervasive HCI offers a definitive resolution to the systemic vulnerabilities of traditional, acoustic-dependent systems.

Evidence synthesized in this review demonstrates that the ``accuracy bottleneck'' has been significantly mitigated by LLMs serving as high-level semantic priors. By functioning as neural error-correction engines, LLMs bridge the domain gap between fragmented articulatory gestures and continuous natural language, achieving performance parity with state-of-the-art acoustic ASR. Consequently, the research frontier has shifted from raw recognition accuracy toward longitudinal robustness and ``pervasive invisibility"-integrating stable, inter-session decoding into everyday wearable form factors like smart glasses and earables.

Despite these milestones, ubiquitous adoption remains obstructed by the ``user-dependency paradox.'' Overcoming anatomical variability through zero-shot transfer learning and self-supervised pre-training remains a fundamental challenge. Furthermore, as decoders approach the realm of imagined intent, the imperative of ``neuro-security'' becomes paramount. Ensuring the inviolability of the boundary between communicative intent and private reflection requires sophisticated cognitive filters and secure, on-device processing architectures to preserve mental agency in an increasingly interfaced world.

Looking ahead, we anticipate a phased adoption cycle, beginning with restorative healthcare and clandestine tactical communication before permeating the broader consumer market. By decoupling human expression from the acoustic constraints of the vocal tract, SSIs are poised to redefine the boundaries of human-computer symbiosis. Ultimately, SSI technology restores the fundamental right of communication where restricted and augments digital agency in a sound-sensitive world, evolving from a prosthetic intervention into a seamless extension of human capability.

\section{Acknowledgments}
This work is supported by National Science and Technology Major Project (2023ZD0121101), National University of Defense Technology (ZZCX-ZZGC-01-04).

\bibliographystyle{IEEEtran}
\bibliography{Bibliography}

\vspace{-5mm}
\begin{IEEEbiography}[{\includegraphics[width=1in,height=1.25in,clip,keepaspectratio]{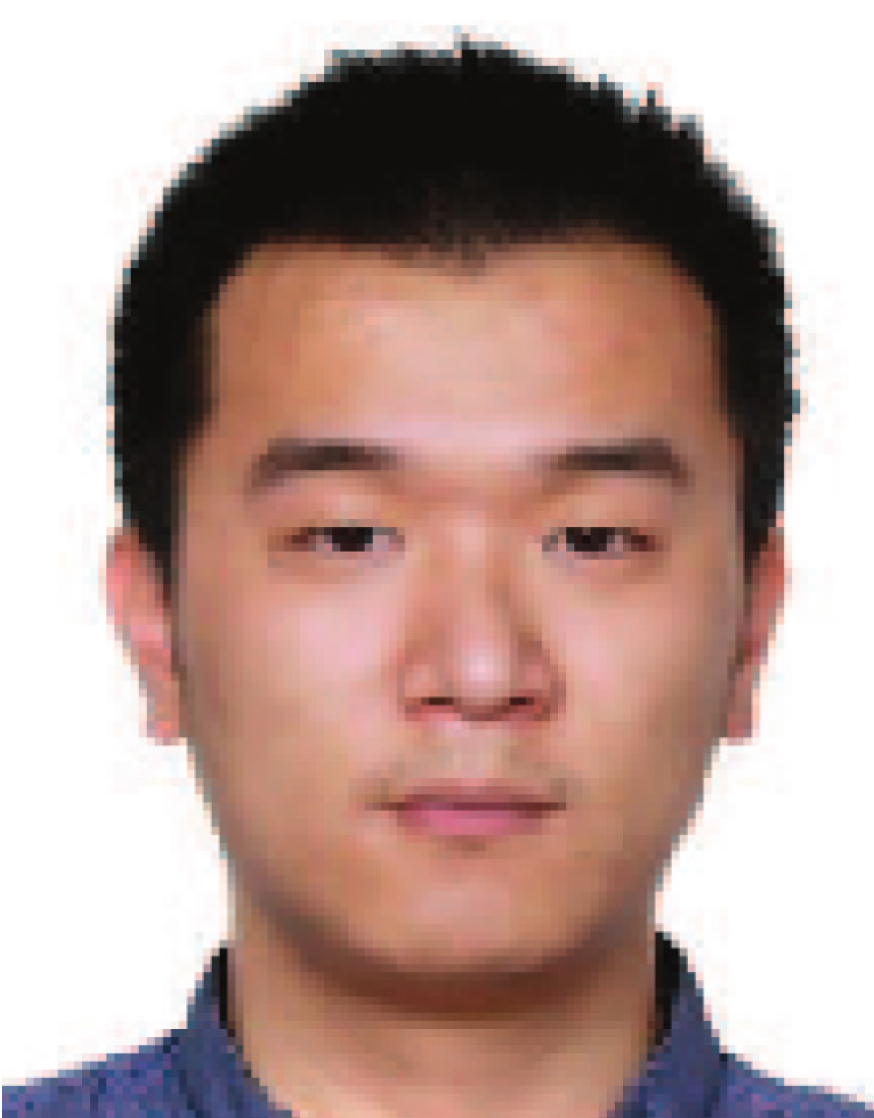}}]{Kele Xu}
(Senior Member, IEEE) received his Doctoral degree from Paris VI University, France, in 2017. He is now an Associate Professor at the School of Computer Science, National University of Defense Technology, China. He has (co-)authored over 100 publications in journals and conferences like ICML, CVPR, NeurIPS, ICLR, AAAI, IJCAI, ASE, and ACM MM. His research focuses on audio signal processing, machine learning, and intelligent software systems.
He also serves as the Associate Editor for IEEE Transactions on Circuits and Systems for Video Technology and the Guest Editor for Science Partner Journal Cyborg and Bionic Systems.
\end{IEEEbiography}

\begin{IEEEbiography}
[{\includegraphics[width=1in,height=1.25in,clip,keepaspectratio]{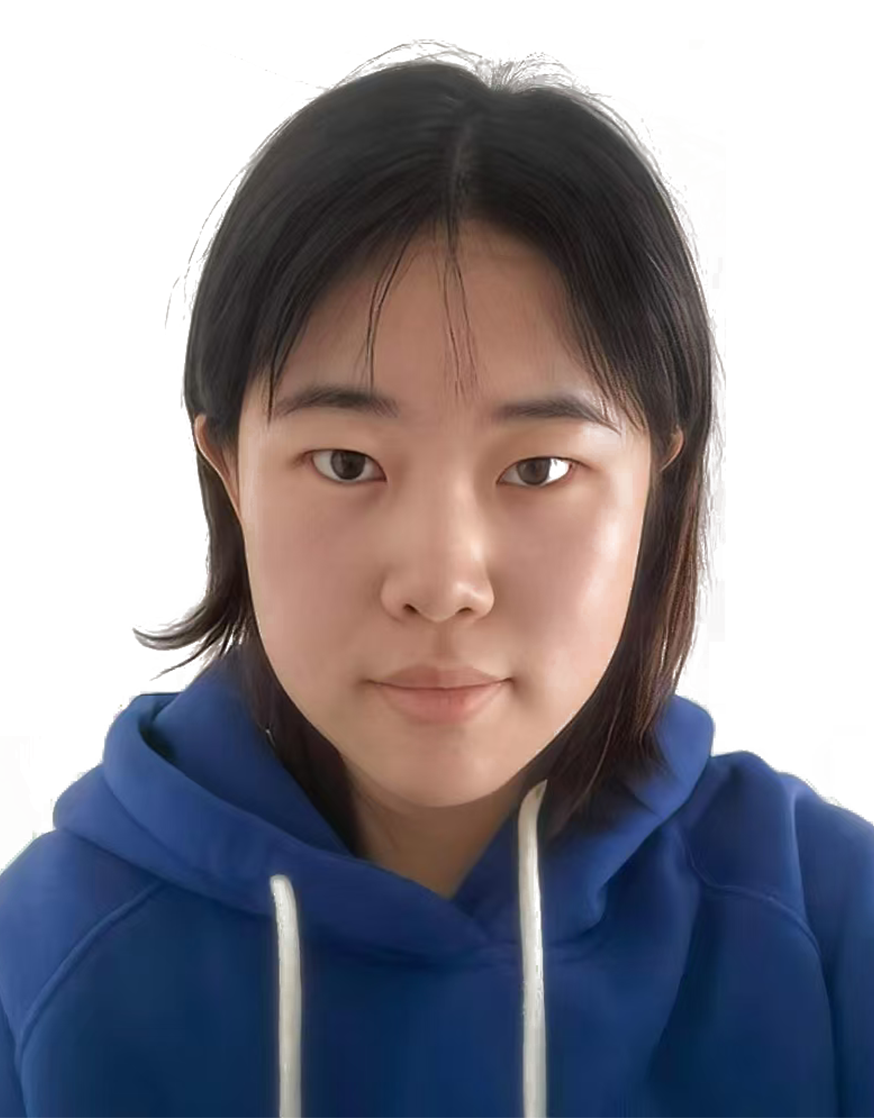}}]{Yifan Wang}
received the B.E. and the Ph.D. degrees from Ocean University of China, Qingdao, China, in 2020 and 2025, respectively. She is currently a postdoctoral researcher  at the School of Computer Science, National University of Defense Technology, Changsha, China. Her research interests include audio signal processing and recognition. Her current research focuses on learning algorithms, active learning, and neural network design.
\end{IEEEbiography}

\begin{IEEEbiography} [{\includegraphics[width=1in,height=1.25in,clip,keepaspectratio]{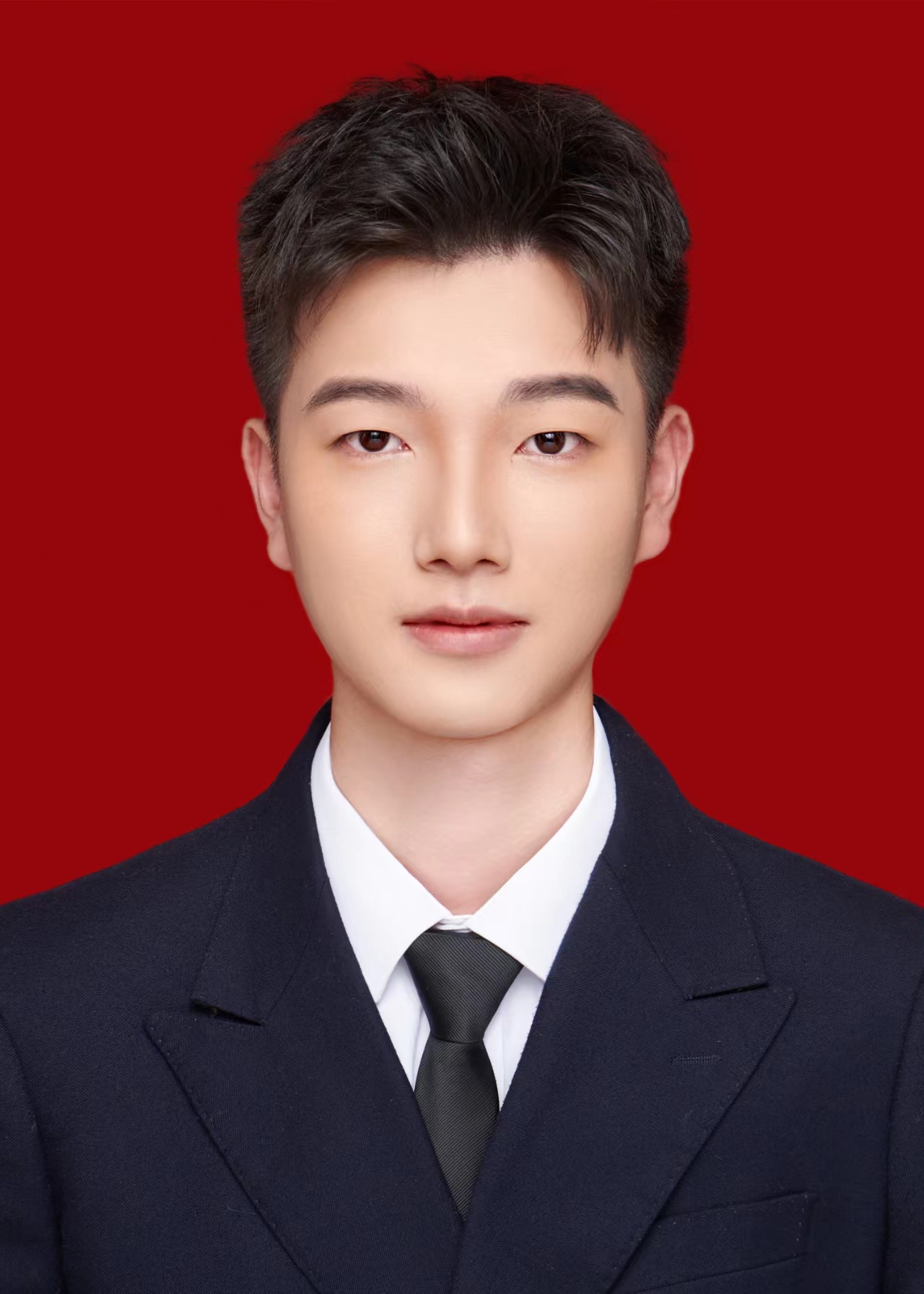}}]{Qisheng Xu}
is a Ph.D. candidate at the College of Computer Science and Technology, National University of Defense Technology. He received his M.S. degree from the National University of Defense Technology in 2024 and his B.S. degree from Wuhan University in 2021. His research interests include audio signal processing, underwater acoustic target recognition, and continual learning algorithms.
\end{IEEEbiography}

\begin{IEEEbiography} [{\includegraphics[width=1in,height=1.25in,clip,keepaspectratio]{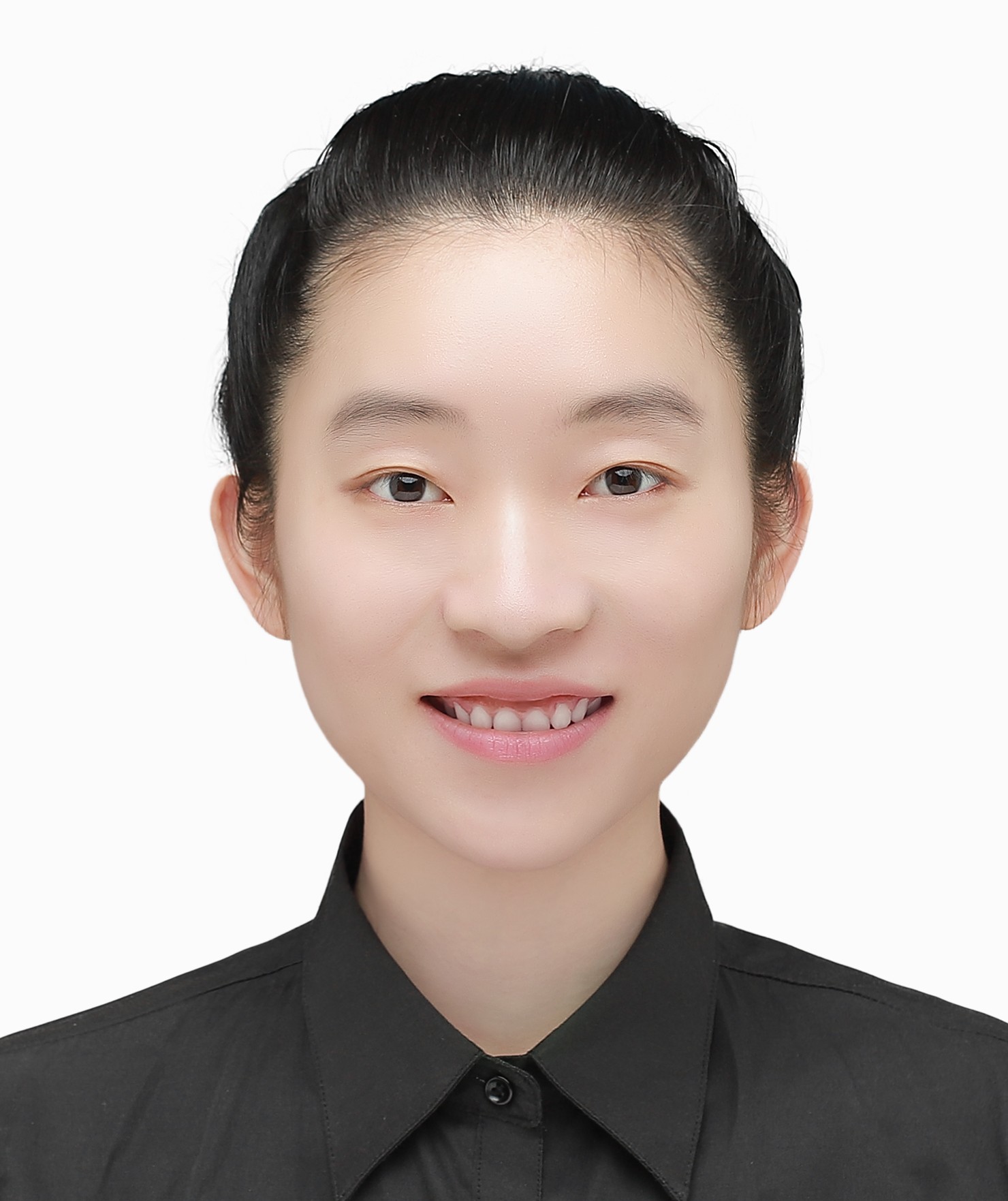}}]{Wuyang Chen}
received the Ph.D. degree in Computer Science and Technology from the National University of Defense Technology, Changsha, China, in 2025. She is currently a Lecturer with the College of Information Science and Engineering, Hunan Normal University, Changsha, China. Her research interests include multimodal learning, voice-face association, and related fields.
\end{IEEEbiography}

\begin{IEEEbiography} [{\includegraphics[width=1in,height=1.25in,clip,keepaspectratio]{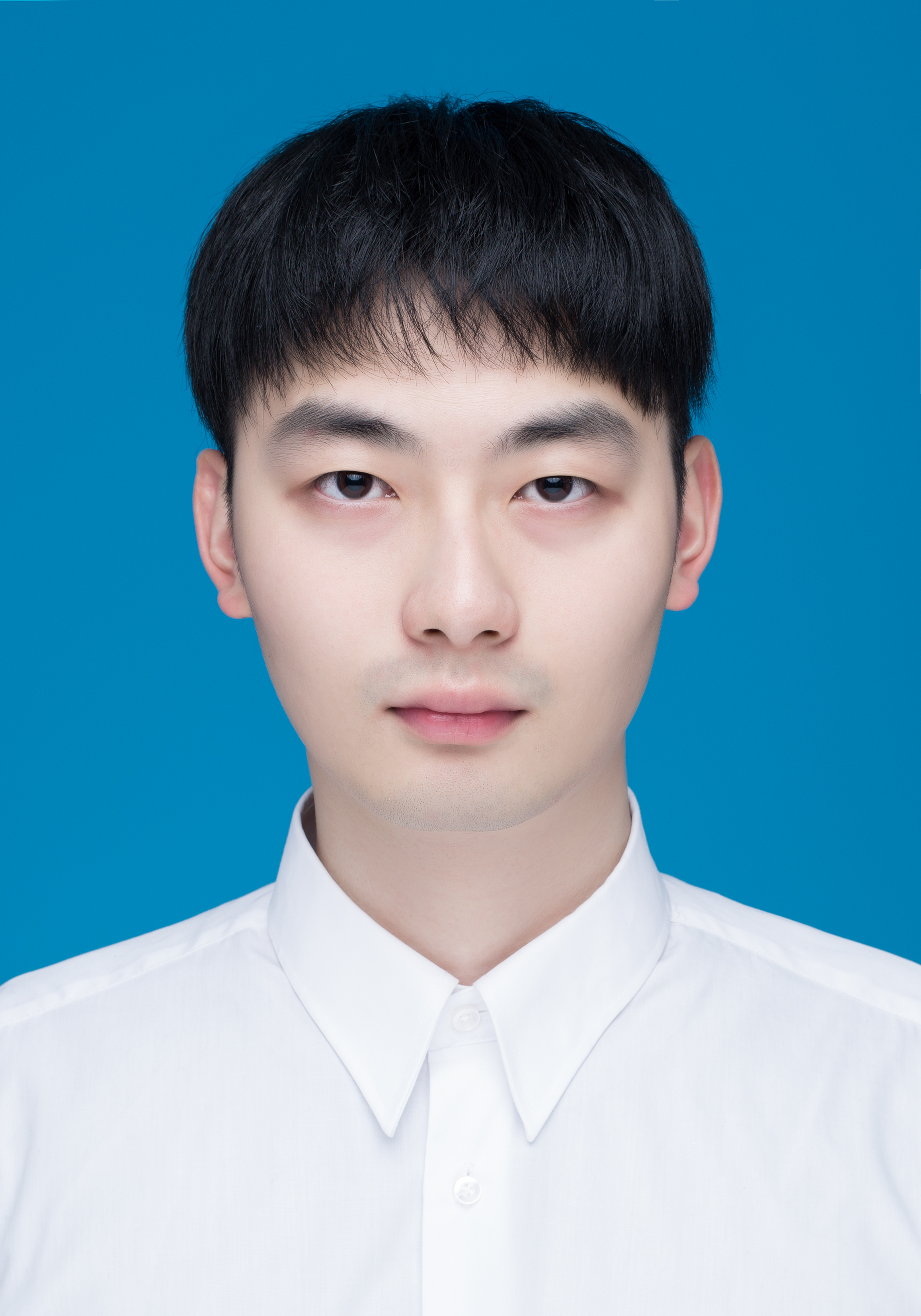}}]{Yutao Dou}
received his B.E. degree in Software Engineering from University of Canberra, Canberra, ACT, Australia, in 2020; and MPhil. degree in University of Sydney, in 2022; and now he is currently a Ph.D. student in the College of Computer Science and Electronic Engineering at Hunan University, Hunan, Changsha, China. His research interests mainly include distributed computing, bioinformatics, and artificial intelligence.
\end{IEEEbiography}

\begin{IEEEbiography} [{\includegraphics[width=1in,height=1.25in,clip,keepaspectratio]{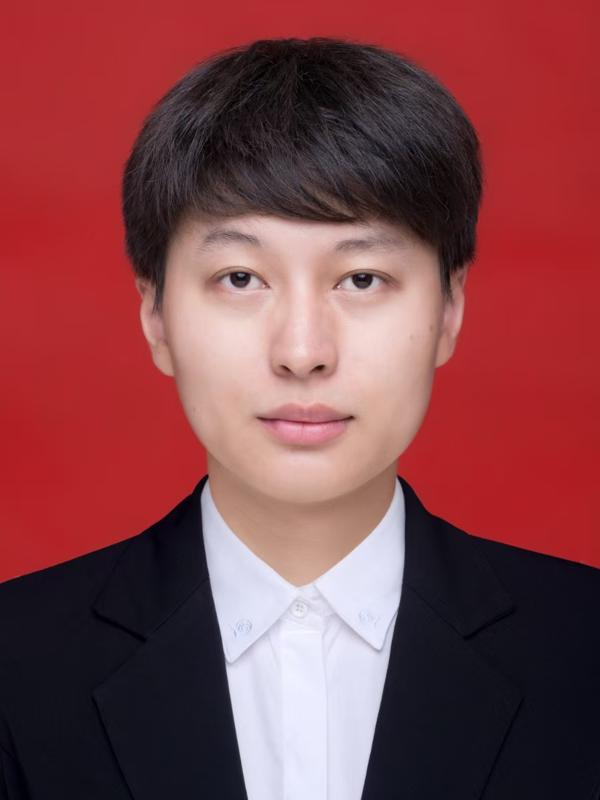}}]{Ming Feng}
received his Ph.D. degree from Tongji University, Shanghai, China, in 2025; and he is currently a postdoctoral researcher in Software Engineering at the National University of Defense Technology, Changsha, China. His research interests mainly include speech processing, medical image processing, weakly supervised learning, handwriting recognition, and related artificial intelligence technologies.
\end{IEEEbiography}

\begin{IEEEbiography}
[{\includegraphics[width=1in,height=1.25in,clip,keepaspectratio]{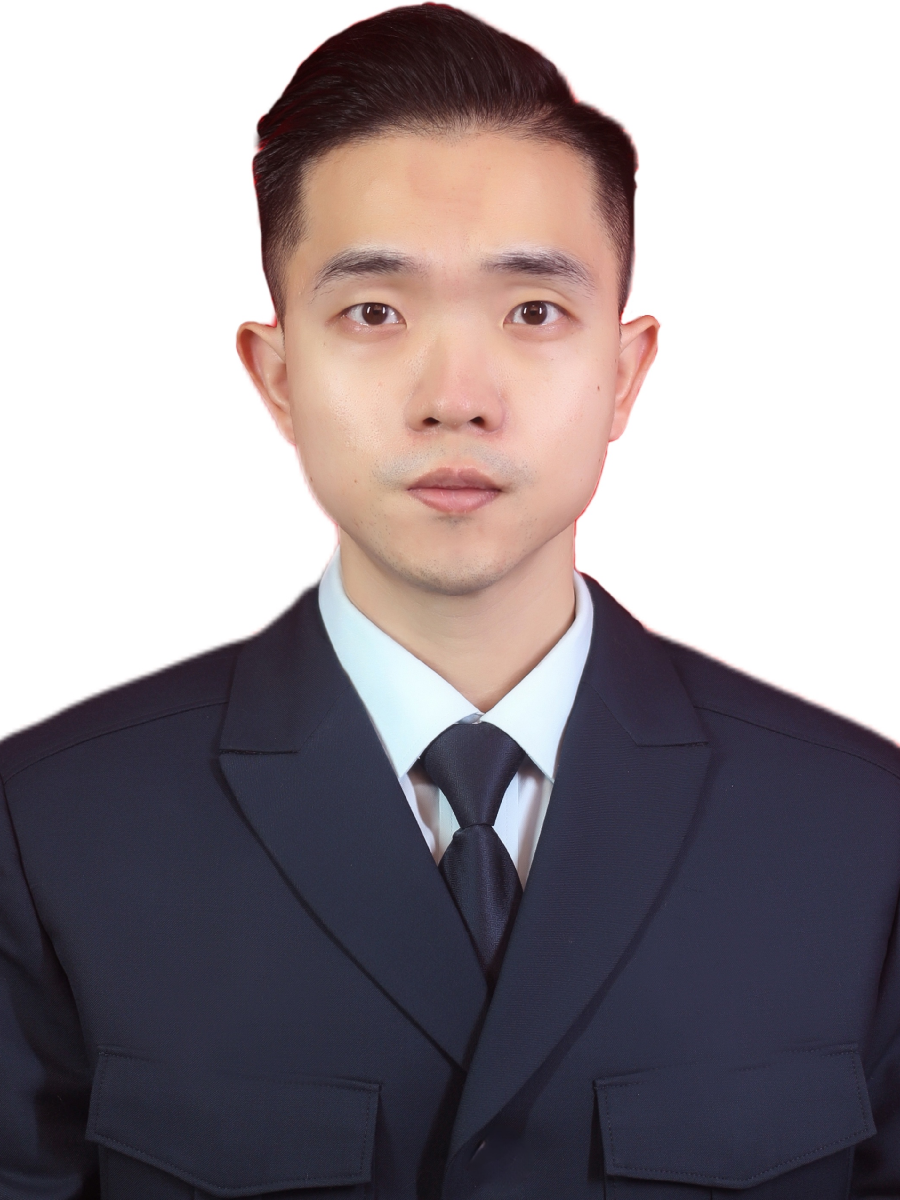}}]{Cheng Yang}
an associate Professor at the National University of Defense Technology (NUDT). He received his Ph.D. in Computer Science from NUDT in 2018. His research interests include audio signal processing and recognition. His current research focuses on learning algorithms, incremental learning, and neural network design.
\end{IEEEbiography}

\vspace{-5mm}
\begin{IEEEbiography}[{\includegraphics[width=1in,height=1.25in,clip,keepaspectratio]{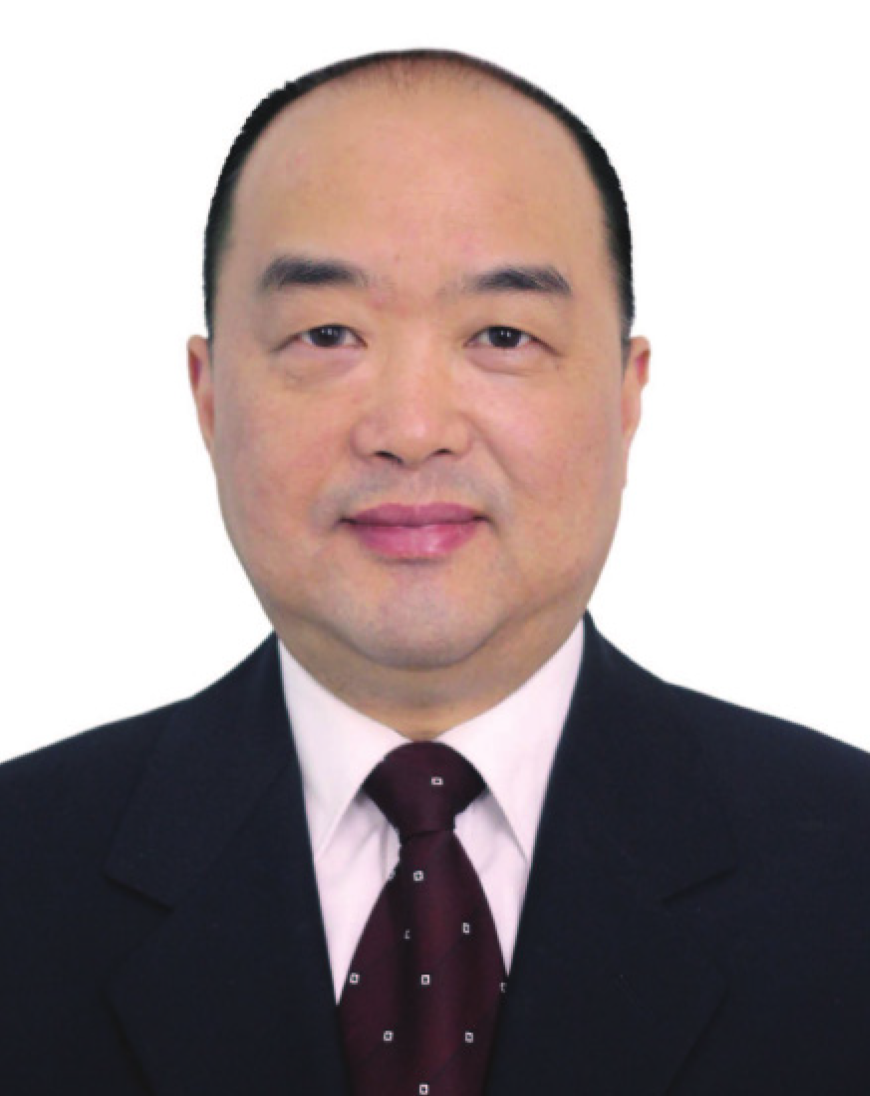}}]{Huaimin Wang} is an Academician of the Chinese Academy of Sciences (CAS) and a Professor at the National University of Defense Technology (NUDT). He received his Ph.D. in Computer Science from NUDT in 1992.
He is also a recipient of the National Science Fund for Distinguished Young Scholars. Dr. Wang has published over 100 research papers in prestigious international conferences and journals. His current research interests focus on middleware, software agents, and trustworthy computing.
\end{IEEEbiography}

\end{document}